\documentclass[twocolumn]{aastex63}
\usepackage{mathrsfs}
\usepackage[]{txfonts}
\usepackage{comment}
\usepackage{multirow}

\shortauthors{Miyakawa et al.}
\shorttitle{Planet Detectability in Young Clusters}
\begin{document}

\title{Color Dependence of the Transit Detectability for Young Active M-dwarfs}

\def\myemail{kohei.miyakawa@nao.ac.jp}
\def\titech{Department of Earth and Planetary Sciences, Tokyo Institute of Technology, Meguro-ku, Tokyo, 152-8551, Japan}
\def\oao{Okayama Astrophysical Observatory, National Astronomical Observatory of Japan, Asakuchi, Okayama 719-0232, Japan}
\def\naoj{National Astronomical Observatory of Japan, NINS, 2-21-1 Osawa, Mitaka, Tokyo 181-8588, Japan}
\def\tokyo{Department of Astronomy, Graduate School of Science, The University of Tokyo, Hongo 7-3-1, Bunkyo-ku, Tokyo, 113-0033, Japan}
\def\komaba{Komaba Institute for Science, The University of Tokyo, 3-8-1 Komaba, Meguro, Tokyo 153-8902, Japan}
\def\Canaria{Instituto de Astrof\'isica de Canarias, V\'ia L\'actea s/n, E-38205 La Laguna, Tenerife, Spain}
\def\abc{Astrobiology Center,  2-21-1 Osawa, Mitaka, Tokyo 181-8588, Japan}
\def\hawaii{Department of Earth Sciences, University of Hawaii at M$\tilde{a}$noa, Honolulu, HI 96822, USA}
\def\norcal{Department of Physics and Astronomy, The University of North Carolina at Chapel Hill, Chapel Hill, NC 27599, USA}

\author{Kohei Miyakawa}\affiliation{\naoj}
\author[0000-0003-3618-7535]{Teruyuki Hirano}\affiliation{\naoj}\affiliation{\abc}
\author{Bun'ei Sato}\affiliation{\titech}
\author{Satoshi Okuzumi}\affiliation{\titech}
\author[0000-0002-5258-6846]{Eric Gaidos}\affiliation{\hawaii}
\begin{abstract}
	We investigate the planetary transit detectability
	in the presence of stellar rotational activity from light curves for young M-dwarfs
	and estimate improvements of the detection at near-infrared (NIR)  wavelengths.
	{Making maps of the transit signal detection efficiency
	over the orbital period and planetary radius} with
	light curves of members of four clusters, Hyades, Praesepe, Pleiades, and Upper Scorpius
	observed by the {\it K2} mission,
	we evaluate the detectability for the rotation period and modulation semi-amplitude.
	{We find that} the detection efficiency remarkably decreases to about 20\%
	for rapidly rotators with $P_{\rm rot} \leq 1$ d
	{ and the lack of planets in Pleiades is likely
	due to the high fraction of rapidly rotating M-dwarfs.}
	We also evaluate the improvements of the planet detection with NIR photometry
	via tests using mock light curves
	assuming that the signal amplitude of stellar rotation decreases at NIR wavelengths.
	Our results suggest that {NIR photometric monitoring would double relative detection efficiency
	for transiting planetary candidates with $P_{\rm rot} \leq 1$ d
	and find planets around M-dwarfs with approximately 100 Myr missing
	in the past transit surveys from the space.}

\end{abstract}
\section{Introduction}\label{sec:intro}
	Although thousands of planets have been confirmed in the last few decades,
	the formation and evolution processes of planetary systems remain veiled in many aspects.
	Planetary evolution can be inferred in snapshots of young planetary systems
	in specific age stages.
	Recently, the secondary mission of
	the {\it Kepler} space telescope {\it K2} \citep{Borucki2010, Howell2014}
	and {\it TESS} \citep{Ricker2015} reported young planets
	belonging to clusters \citep[e.g.,][]{Mann2016, Mann2016b, Newton2019, Rizzuto2020},
	which have provided insights into the physics of the young systems.
	For example, radial inflation of young planets likely leads to atmospheric escape
	through stellar radiation \citep{Owen2019}.
	The number of detected planets in each cluster
	may also constrain the time-scale of their evolutionary stage.

	Planetary systems around low luminosity M-dwarfs
	are targeted for biosignatures,
	because their habitable planet(s) might locate close to the host star,
	which is observationally preferable \citep{Scalo2007, Kopparapu2013}.
	Moreover, the observable planetary signals
	in both radial-velocity (RV) measurements and transit photometry
	are larger than those around solar-type stars, implying that
	{M-dwarfs are ideal targets for finding small rocky planets and studying
	such planets' atmospheres \citep[e.g.,][]{Burke2014}.}
	\begin{table*}
		\centering
		\caption{Young transiting planets confirmed in clusters or associations
		by the space missions.}\label{tab:planets}
		\begin{tabular}{lcccc}
		\hline \hline
		Name & Host star temperature [K] &Planetary radius [$R_{\oplus}$] & Orbital period [d]& Reference \\
		\hline
		\multicolumn{5}{c}{$Upper~Sco ~~(\sim 10 ~{\rm Myr})$} \\
		K2-33 b $^*$ & $3540\pm70$ & $5.04_{-0.37}^{+0.34}$ & $5.425 \pm<0.001$
		&  \citet{David2016}; \citet{Mann2016b}$^\circ$\\
		\hline
		\multicolumn{5}{c}{$Low~Cen ~~(\sim 11 ~{\rm Myr})$} \\
		TOI-1227 b & $3072\pm74$  & $9.37_{-0.57}^{+0.74}$ & $27.363 \pm<0.001 $&  \citet{Mann2021}\\
		\hline
		\multicolumn{5}{c}{$Sco{\rm-}Cen ~~(\sim 17 ~{\rm Myr})$} \\
		HIP67522 b & $5675\pm75$ & $9.72_{-0.47}^{+0.48}$ & $6.960 \pm<0.001$ &  \citet{Rizzuto2020}\\
		\hline
		\multicolumn{5}{c}{$Beta~Pictoris~Moving~Group ~~(\sim 22 ~{\rm Myr})$} \\
		AU Mic b & $3700\pm100$ & $4.20\pm0.20$ & $8.463 \pm <0.001$ &  \citet{Plavchan2020}\\ \cline{5-5}
		AU Mic c & $3700\pm100$ & $2.79_{-0.30}^{+0.31}$ & $18.859 \pm <0.001$&  \citet{Martioli2021}; \citet{GIlbert2022}$^\circ$\\
		\hline
		\multicolumn{5}{c}{$Taurus{\rm-}Auriga ~~(\sim 23 ~{\rm Myr})$} \\
		V1298 Tau b & $4970\pm120$ & $10.27_{-0.53}^{+0.58}$ & $24.140 \pm 0.002$ & \multirow{4}{*}{\citet{David2019}}\\
		V1298 Tau c & $4970\pm120$ & $5.59_{-0.32}^{+0.36}$ & $8.250 \pm <0.001$ & \\
		V1298 Tau d & $4970\pm120$ & $6.41_{-0.40}^{+0.45}$ & $12.403 \pm 0.002$ &  \\
		V1298 Tau e & $4970\pm120$ & $8.74_{-0.72}^{+0.84}$ & $60_{-18}^{+60}$ &  \\
		\hline
		\multicolumn{5}{c}{$IC~2602 ~~(\sim 35 ~{\rm Myr})$} \\
		TOI-837 b & $6407\pm162$  & $8.45\pm{0.99}$ & $8.325 \pm <0.001$ &  \citet{Bouma2020}\\
		\hline
		\multicolumn{5}{c}{$Tuc{\rm - }Hor ~~(\sim 40 ~{\rm Myr})$} \\
		DS Tuc A b & $5542\pm21$  & $5.63_{-0.21}^{+0.22}$ & $8.139 \pm0.001$ &  \citet{Benatti2019}$^\circ$; \citet{Newton2019}\\
		\hline
		\multicolumn{5}{c}{$Cep{\rm - }Her ~~(\sim 40 ~{\rm Myr})$} \\
		Kepler-1627 A b& $5505\pm60$  & $3.78\pm{0.16}$ & $7.203 \pm <0.001$ &  \citet{Bouma2022a}\\ \cline{5-5}
		Kepler-1643 b& $4916\pm110$  & $2.32\pm{0.14}$ & $5.323 \pm <0.001$ &  \citet{Bouma2022b}\\ \cline{5-5}
		KOI-7368 b& $5241\pm 100$  & $2.22\pm{0.12}$ & $6.843 \pm <0.001$ &  \citet{Bouma2022b}\\ \cline{5-5}
		KOI-7913 A b& $4324\pm70$  & $2.34\pm{0.18}$ & $24.279 \pm <0.001$ &  \citet{Bouma2022b}\\
		\hline
		\multicolumn{5}{c}{$MELANGE-3 ~~(\sim 105 ~{\rm Myr})$} \\
		Kepler-1928 b & $5720\pm60$ & $1.96_{-0.04}^{+0.06}$ & $19.578 \pm <0.001$ & \citet{Barber2022}\\ \cline{5-5}
		Kepler-970 b &	$4290\pm70$ & $2.47_{-0.06}^{+0.11}$ & $16.737 \pm <0.001$ & \citet{Barber2022}\\
		\hline
		\multicolumn{5}{c}{$Psc-Eri~~(\sim 120 ~{\rm Myr})$} \\
		TOI-451 b & $5550\pm56$  & $1.91\pm{0.12}$ & $1.859 \pm<0.001$ &  \multirow{3}{*}{\citet{Newton2021}}\\
		TOI-451 c & $5550\pm56$  & $3.1\pm{0.13}$ & $9.193 \pm<0.001$ & \\
		TOI-451 d & $5550\pm56$  & $4.07\pm{0.15}$ & $16.365 \pm<0.001$ &  \\
		\hline
		\end{tabular}
	\end{table*}
	\begin{table*}
		\centering
		Continuation of Table \ref{tab:planets}\\
		\begin{tabular}{lcccc}

		\hline
		Name & Host star temperature [K] &Planetary radius [$R_{\oplus}$] & Orbital period [d]& Reference
		\\ \hline

		\multicolumn{5}{c}{a co-moving star group ~~$(\sim  200~{\rm Myr})$} \\
		TOI-1807 b & $4757_{-50}^{+51}$  & $1.85\pm{0.04}$ & $0.549 \pm<0.001$ &  \citet{Hedges2021}\\ \cline{5-5}
		TOI-2076 b & $5187_{-53}^{+54}$  & $3.28\pm{0.04}$ & $10.356 \pm<0.001$ &  \multirow{3}{*}{\citet{Hedges2021}}\\
		TOI-2076 c & $5187_{-53}^{+54}$  & $4.44\pm{0.05}$ & ... &  \\
		TOI-2076 d & $5187_{-53}^{+54}$  & $4.14\pm{0.07}$ & ... &  \\

		\hline
		\multicolumn{5}{c}{$Group-X ~~(\sim  300~{\rm Myr})$} \\
		TOI-2048 b & $5185\pm60$ & $2.05_{-0.19}^{+0.20}$ & $13.790\pm0.001$ & \citet{Newton2022}\\
		\hline
		\multicolumn{5}{c}{$Ursa~Major~~(\sim 400 ~{\rm Myr})$} \\
		HD 63433 b & $5640\pm74$  & $2.15\pm0.10$ & $7.108 \pm <0.001$ &  \multirow{2}{*}{\citet{Mann2020}}\\
		HD 63433 c & $5640\pm74$  & $2.67\pm0.12$ & $20.545 \pm 0.001$ &  \\
		\hline
		\multicolumn{5}{c}{$Praesepe~~(\sim 700 ~{\rm Myr})$} \\
		K2-95 b $^*$ & $3410\pm65$  & $3.7\pm0.2$ & $10.135\pm0.001$ &  \citet{Obermeier2016}; \citet{ Mann2017}$^\circ$\\  \cline{5-5}
		K2-100 b & $6120\pm90$  & $3.5\pm0.2$ & $1.674\pm<0.001$ &  \citet{Mann2017}\\  \cline{5-5}
		K2-101 b & $4819\pm45$  & $2.0\pm0.1$ & $14.677\pm0.001$ &  \citet{Mann2017}\\ \cline{5-5}
		K2-102 b & $4695\pm50$  & $1.3\pm0.1$ & $9.916\pm0.001$ &  \citet{Mann2017}\\ \cline{5-5}
		K2-104 b $^*$ & $3660\pm67$  & $1.9_{-0.1}^{+0.2}$ & $1.974\pm0.001$ &  \citet{Mann2017}\\ \cline{5-5}
		K2-264 b $^*$ & $3580\pm70$  & $2.27_{-0.16}^{+0.20}$ & $5.840\pm<0.001$ & \multirow{2}{*}{\citet{Rizzuto2018}}\\
		K2-264 c & $3580\pm70$  & $2.77_{-0.18}^{+0.20}$ & $19.664\pm<0.001$ & \\
		\hline
		\multicolumn{5}{c}{$Hyades~~(\sim 700 ~{\rm Myr})$} \\
		HD 283869 b & $4655\pm55$  & $1.96\pm0.13$ & $106_{-25}^{+74}$ &  \citet{Vanderburg2018}\\ \cline{5-5}
		K2-25 b $^*$ & $3180\pm60$  & $3.31_{-0.25}^{+0.34}$ & $3.485\pm<0.001$ &  \citet{Mann2016}\\ \cline{5-5}
		K2-136A b & $4499\pm50$  & $0.99_{-0.04}^{+0.06}$ & $7.975\pm0.001$ &
		\multirow{3}{*}{\begin{tabular}{c} \citet{Ciardi2018, Livingston2018};\\ \citet{Mann2018}$^\circ$
		\end{tabular}}\\
		K2-136A c & $4499\pm50$  & $2.91_{-0.10}^{+0.11}$ & $17.307\pm<0.001$ & \\
		K2-136A d & $4499\pm50$  & $1.45_{-0.08}^{+0.11}$ & $25.575\pm0.002$ & \\
		\hline
		\multicolumn{5}{c}{$NGC 6811~~(\sim 900 ~{\rm Myr})$} \\
		Kepler-66 b & $5962\pm79$  & $2.80\pm0.16$ & $17.816\pm<0.001$ &  \citet{Meibom2013}\\ \cline{5-5}
		Kepler-67 b & $5331\pm63$  & $2.94\pm0.16$ & $15.726\pm<0.001$ &  \citet{Meibom2013}\\
		\hline
		\multicolumn{5}{c}{$ Ruprecht~147 ~~(\sim 3 ~{\rm Gyr})$} \\
		K2-231 b & $5695\pm50$  & $2.5\pm0.2$ & $13.842\pm0.001$ &  \citet{Curtis2018}\\
		\hline \hline

		\end{tabular}
		\begin{flushleft}
		{\bf Note}:  The targets marked with asterisk are systems
		with effective temperature of 3000 K - 4000K
		and orbital period of $\leq$ 10 d.
		We cite the orbital parameters from the references marked with open circle
		when two or more papers exist for each system.
		\end{flushleft}
	\end{table*}

	However, young stars usually have highly active regions on their surface
	(e.g., star spots, plage, and faculae) and high rotation velocities,
	which manifest as significant variations on the light curves of space missions \citep{Rizzuto2017}.
	This variation contaminates the light curve and obfuscates the planetary transit signals.
	{ Stellar rotational activities also appear as noise or false planetary signals
	(so-called ``stellar jitter") in RV measurements \citep{Klein2020, Cale2021, Klein2022}. }
	Consequently, only approximately 30 transiting planetary systems,
	which are shown in Table \ref{tab:planets},
	have been confirmed and/or validated around young stars in the clusters.
	For M-dwarfs, whose effective temperatures are 3000 - 4000 K, only 8 systems are confirmed.


	In order to constrain the time scales of planetary formation and migration,
	one must estimate the true frequency of young planets.
	In fact,  the distribution of the currently discovered planets appears to be biased
	with respect to stellar ages.
	For example, planets have been detected in the very young Upper Scorpius cluster ($\sim 8$ Myr) \citep[e.g., ][]{Mann2016b}
	and in Hyades and Praesepe ($\sim 700$ Myr) \citep[e.g.,][]{Mann2016, Rizzuto2018},
	where evolution events are assumedly completing,
	but no planets have been found in Pleiades, whose age is intermediate between Upper Scorpius
	and Hyades ($\sim 125$ Myr).
	Additionally, no planetary systems are detected around M-dwarfs with the age of about 100 Myr
	for all targeted clusters as listed in Table \ref{tab:planets}.
	If the planetary systems both outside and inside of clusters
	share unique statistical properties,
	their age distribution should indicate the evolutionary process itself.
	However, whether this trend is caused by true planetary frequency
	or the detection efficiency of the transit signal is unclear,
	{ because the decline of the planet detection might be affected by stellar activity.
	The stellar rotation period varies in the early stage
	due to spin-up with contraction and constant angular momentum
	and spin-down with braking by the magnetic field \citep{Godoy-Rivera2021}.}


	In analyses of {\it TESS} photometry,
	\citet{Nardiello2021} suggested that the detection biases in clusters depend on the photometric precision.
	We hypothesize that the planetary transit detectability around young active stars in {\it K2}
	is significantly prevented by photometric variations in the light curves.
	{The detectability with the stellar activity was often discussed for each cluster
	in previous studies  \citep[e.g.,][]{Mann2018},
	but the systematic interpretation across the clusters is not enough
	{ to conclude whether current planet distribution is biased by stellar activity or not.}}

	{ Tansit detection of young planets will be improved at NIR wavelengths,
	because the flux variation by stellar activity, which has hindered transit detection,
	is wavelength dependent.}
	Especially for M-dwarfs, the photometric variation amplitudes are expected to be much lower
	at NIR wavelengths.
	{Recent NIR photometric observations have gradually confirmed this wavelength dependence
	for some young stars \citep{Frasca2009, Morris2018, Miyakawa2021b}.}
	{ For example, \citet{Miyakawa2021b} suggest the variation amplitude might
	decreases from 56\% to 17\% in $J$ - band compared to in $Kp$ - band.}
	If future NIR observations with either space and/or ground-based telescopes are carried out,
	the planetary systems around young stars overlooked in previous space missions
	performed at visible wavelengths might be discovered.

	In this study, {we evaluate the planetary detection efficiencies around M-dwarfs
	with respect to stellar rotation }
	to understand the causes of the current planet distribution in stellar clusters.
	We perform {calculations of the signal detection efficiency (SDE)
	over the orbital period and signal amplitude
	through the detection tests with the {\it K2} light curves}
	{ and discuss the systematic detectability across four clusters}:
	Hyades, Praesepe, Pleiades, and Upper Scorpius.
	Furthermore, transit detection efficiencies are derived
	as a benchmark for the statistical assessment of young planets.
	We also estimate the extent to which observations at the NIR wavelengths
	improves the detectability and assess the benefit of these improvements for future observational missions.

	\begin{figure*}[ht]
		\centering
		\includegraphics[width=18cm]{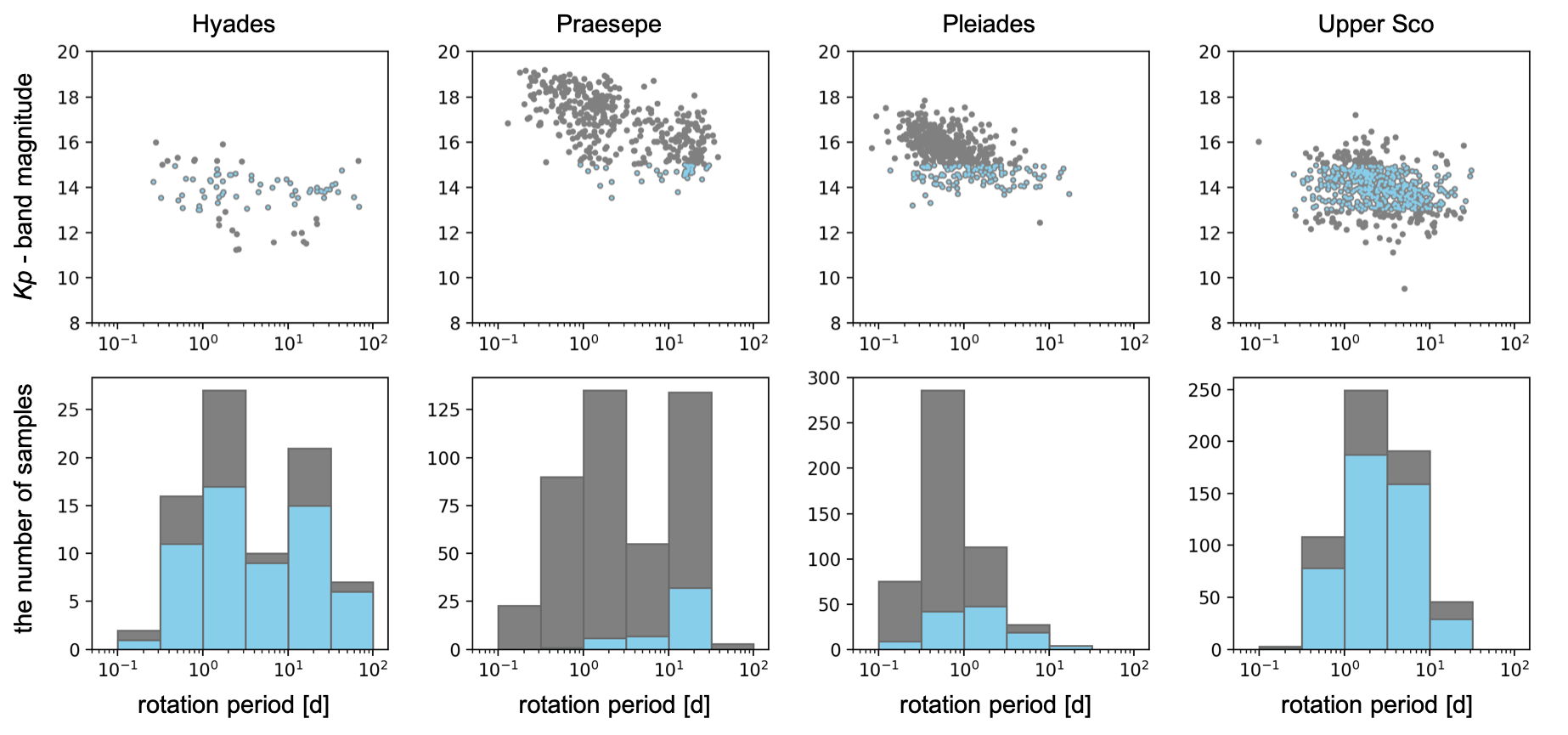}
		\caption{Top panels: scatter plots of the rotation period vs. apparent {\it Kp} magnitude
		for { stars with 3000 K - 4000 K} in four clusters.
		Bottom panels: histograms of the rotation periods in the clusters.
		The gray and blue data indicate the original members from the literatures
		and selected targets in this study, respectively.}
		\label{pic:hist}
	\end{figure*}



	The remainder of this paper is organized as follows.
	Section \ref{sec:target} introduces our targets
	and { their selection process}.
	{Section \ref{sec:method} details the SDE mapping through the detection tests of transiting planets
	around active stars} 
	and Section \ref{sec:res} shows the test results over the clusters.
	Moreover, Section \ref{sec:nir} shows improvement of the detection efficiency in the NIR
	through the test with mock light curves.
	Section \ref{sec:discussion} discusses biased frequencies of planets around active stars
	based on current observational results.
	A summary is given in Section \ref{sec:sum}.

\section{Target Selection}\label{sec:target}
	This study focuses on planetary systems around young rapidly rotating M-dwarfs.
	The statistical properties of stars belonging to a given cluster
	are well determined.
	From the young stars in stellar clusters observed in the {\it K2} mission,
	we selected targets whose rotation periods were determined in previous researches.
	Our original target lists { and rotation periods}
	in Pleiades, Praesepe, Upper Scorpius, and Hyades were
	taken from \citet{Rebull2016}, \citet{Douglas2017}, \citet{Rebull2018}, and \citet{Douglas2019}.

	After matching the EPIC ID and 2MASS names of these targets,
	we downloaded their parallax and color information from
	the $Gaia$ EDR3 archive website\footnote{https://gea.esac.esa.int/archive/} \citep{Gaia2016, Gaia2020, GaiaEDR}.
	Because we focus on M-dwarfs,  
	the targets were filtered through the range 3000 - 4000 K
	{based on samples treated in \citet{Mann2015}.}
	The effective temperatures $T_{\rm eff }$ of each target was determined using the empirical
	$B_P - R_P$ vs. $T_{\rm eff}$ relation described in \citet{Mann2015}.
	The apparent magnitudes in  the {\it Kp}- band of the {\it Kepler}
	may also directly affect the detection result.
	\citet{Nardiello2021} showed that the detection efficiency of samples analyzed
	in the PATHOS \citep{Nardiello2020, Nardiello2021} from the {\it TESS} targets.
	improves for brighter targets in the range of 6 - 18 ${\rm mag}_T$.
	To avoid this systematic bias, we filtered the targets through the magnitude range of
	$Kp$ = 13 - 15, sufficiently narrow to omit the trend
	while allowing a balanced representative of the samples in each cluster. 
	After filtering, we obtained 61, 61, 121, and 476 light curves in Hyades, Praesepe, Pleiades,
	and Upper Scorpius, respectively.

	The distributions for the stellar rotation period are shown in Figure \ref{pic:hist}.
	Upper panels show scatter plots of the rotation period vs. apparent magnitudes
	and lower panels show histograms for the rotation period for each cluster.
	The gray and blue data are original samples in the literature
	and selected samples in this study, respectively.
	For reference, the medians of the rotation periods from the literature
	{ are $2.45_{-1.69}^{+22.19}$ d, $2.06_{-1.47}^{+16.49}$ d,
	$0.63_{-0.33}^{+1.19}$ d,  and $2.32_{-1.52}^{+5.09}$ d }
	for Hyades, Praesepe, Pleiades, and Upper Scorpius, respectively.
	{ The uncertainties are estimated as range of 86th - 50th and 50th -14th percentile of the distribution.}
	{For Hyades and Upper Scorpius, we select samples at similar rates from the rotation-period bins.}
	{For Praesepe and Pleiades, relatively slow rotators are selected from the total samples}.
	{Note that we do not filter based on quality flags of the rotation period in the literature,
	thus periods larger than 30 d can be mis-detections or double of actual rotation periods.}


\section{Method}\label{sec:method}
	{ We perform the injection-recovery test of transiting planets for the clusters
	to evaluate the detection biases due to stellar activity (i.e., rotation period and flux variation).
	Firstly, we use light curves observed in {\it K2} mission
	and investigate the biases to constrain true planet frequency for the current survey
	in Section \ref{sec:method} and \ref{sec:res}.
	Moreover, we propose improvements in the planet detection in the NIR
	assuming future planet research in Section \ref{sec:nir}.}

	\subsection{Light Curves}
		We use the light curves extracted
		by the pipeline of \citet[\texttt{K2SFF};][]{Vanderburg2014},
		which are corrected with the centroid position of point-spread function.
		The correction is simply performed by fitting a polynomial function
		with respect to the spacecraft motion.
		The \texttt{K2SFF} light curves of our targets were downloaded
		from the Miltsuki Archive for Space Telescope portal
		website\footnote{https://archive.stsci.edu/k2}.
		{Note that, we avoid using light curves extracted with machine learning - based pipelines,
		because miscorrection may be caused by high-cadence photometric variation and/or
		the existence of contamination of nearby stars in the clusters regions \citep{Luger2016}.}

		%


	\subsection{Injection of Planetary Transit Signals}\label{sec:injection}
		To evaluate the detection efficiency of planetary transits,
		we injected mock signals into the $K2$ light curves over a wide range of orbital parameters.
		To remove the long-term trends from the light curves,
		we first binned the data points into five median bins and subtracted them
		from the original light curves by spline interpolation.
		In the light curves with discontinuous points caused by the spacecraft motion,
		the points before and after the missing points were corrected independently
		such that the light curves were smoothly connected.
		To ensure that the detectability of stellar rotational activities
		was unaffected by anomalies such as flares, the positive
		and negative outliers larger than $4 - \sigma$ were replaced by unity.
		The robust standard deviation $\sigma$ was derived as 1.48 times
		the MAD in the robust statistics.

		Next, we injected the planetary transit signal with the model of \citet{Ohta2009}
		for a given radius ratio $R_{\rm p}/R_{\rm s}$ and orbital period $P_{\rm orb}$.
		The limb-darkening of the host star was modeled with
		the quadratic functions in \citet{Claret2011}.
		To constrain the orbital parameters, the radius and mass were derived using
		the relations in $K$ band $M_K$ in \citet{Mann2015}
		and \citet{Mann2019}, respectively.
		These relations use the absolute magnitude $M_K$ in the $K$ band, which
		is calculated from the parallaxes in $Gaia$ EDR3
		and the $K$ band magnitudes in the 2MASS catalog \citep{2MASS}.
		{ We discuss uncertainties in the estimations of radius
		and effective temperature in Appendix \ref{ap:iso}.}
		To simplify the discussion,
		we set the eccentricity and impact parameter to 0.

		To investigate the detectability in the orbital parameter space for each star,
		we calculated the signal detection efficiency (SDE) over the parameter space.
		We reproduced 100 transit signals on 10 $\times$ 10 grids of $R_{\rm p}/R_{\rm s}$
		versus $P_{\rm orb}$ for each target
		{considering the computational cost.}
		The $R_{\rm p}/R_{\rm s}$ and $P_{\rm orb}$ scales were linearly spaced in [0.01, 0.2]
		and log-linearly spaced in [1 d, 30 d], respectively.
		{The validity of the 10 $\times$ 10 simulations is discussed in Appendix \ref{ap:3030}
		compared with 30 $\times $ 30 simulations.}

	\begin{figure*}[t]
		\centering
		\includegraphics[width=18.5cm]{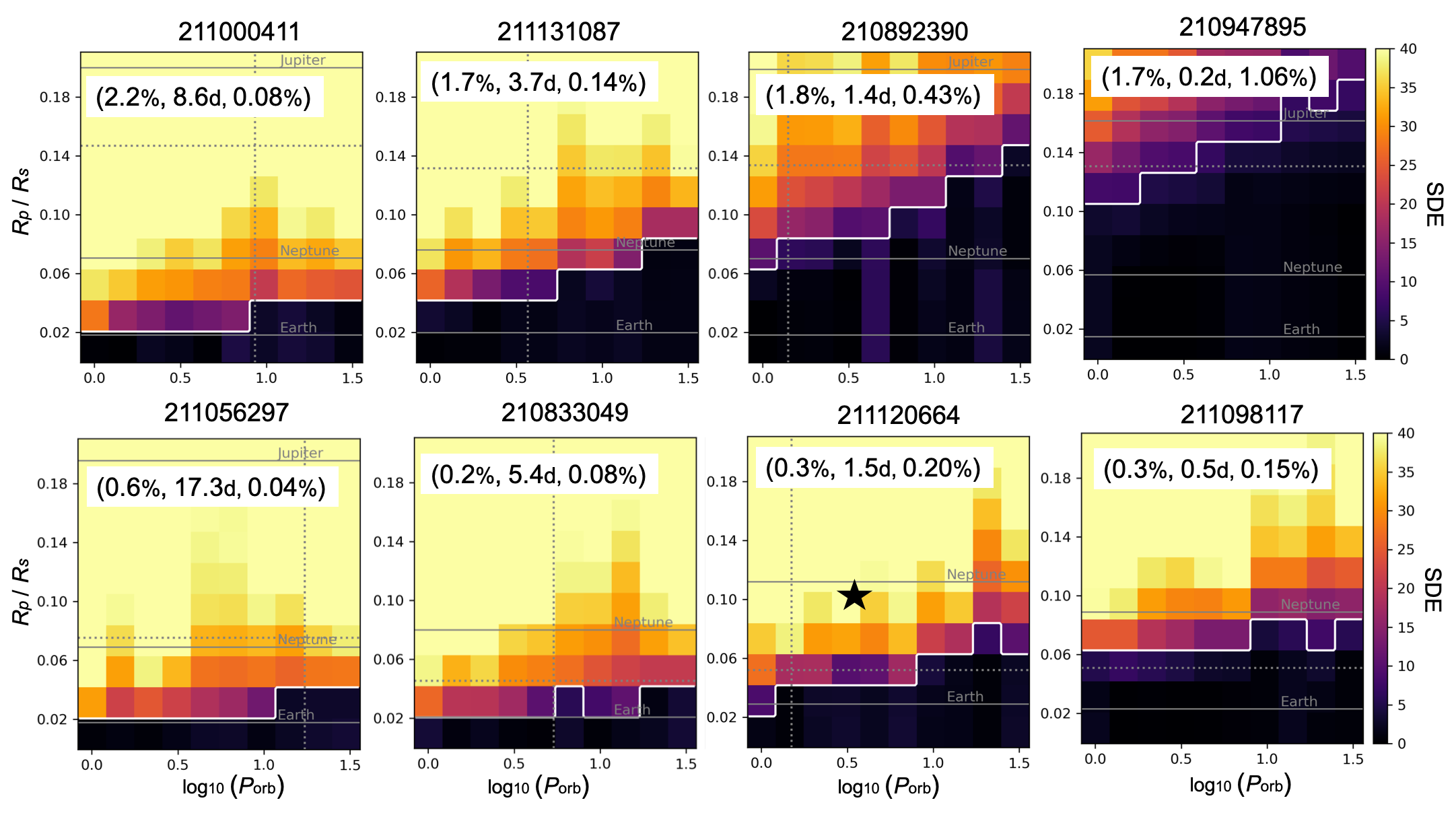}
		\caption{{Examples of heat maps of the transit signal detection efficiency (SDE) for the 8 targets}.
		The colors corresponds to the SDE in the transit least square \citep[TLS; ][]{Hippke2019} test
		with respect to the log-scale orbital period $P_{\rm orb}$
		and relative planetary radius $R_{\rm p}/R_{\rm s}$ of the injected planet.
		The horizontal lines are converted planet radii of Earth, Neptune, or Jupiter,
		included as guides.
		The vertical and horizontal dotted lines represent the rotational period
		and the square root of the flux variation semi-amplitude, respectively.
		The white line is the boundary of SDE = 7 (the detection threshold).
		The photometric variation semi-amplitude, rotation period, { and mean of the standard deviation
		in 0.3 d - sized bins of the detrended light curves
		($A$, $P_{\rm rot}$, $\sigma_{\rm 0.3 d}$) of the targets}
		are shown in the top of the figures.
		{ The black star on the EPIC 211120664 is a projection of K2-25 b
		whose host star is the fastest rotator in confirmed planetary systems
		and shows similar light curve modulation to EPIC 211120664.}
		}
		\label{pic:cmap}
	\end{figure*}

	\subsection{Evaluation of the SDE}\label{sec:recovery}
		{We calculate the SDE of the injected planetary signals
		to evaluate the parameter regions where planets are detectable for each star.
		The normalization of the light curves were performed following the procedure in \citet{Mann2016}.}
		We first computed the Lomb - Scargle (LS) periodogram \citep{Scargle1982}
		to detect the significant variations { which are possibly due to stellar rotational activity.}
		{ We removed these signals in frequency space using the fast Fourier transform (FFT)
		and derived filtered light curves by inverting the spectral data to time space.}
		{ The period search was performed in the range between one and 70 days to retain
		the planetary transit signals.}
		The systematic trend were then removed thorough a median filter with a one-day window,
		and positive excursions larger than 3 - $\sigma$ with $1.48 \times$ MAD
		were replaced by the median value.

		Finally, the transit signals were recovered by
		the transit-least squares \citep[TLS; ][]{Hippke2019, Hippke2019b}  algorithm,
		{which is likely a more effective and reliable method to detect small transit signals than
		a previously popular box-fitting least square algorithm \citep{Kovacs2002}.}
		{Besides, the TLS is optimal for V-shape light curves
		such as close-in planets around small host stars.}
		Here we employed the TLS package implemented
		in \texttt{Python} code \footnote{https://transitleastsquares.readthedocs.io/en/latest/}.
		As the input parameters of the TLS algorithm,
		we applied the stellar mass and radius derived
		by the relations with respect to $M_K$ in \citet{Mann2015}
		and \citet{Mann2019}, respectively.
		{The SDE value is calculated through the TLS
		for each period-radius segment of the 10 $\times$ 10 parameter space.
		To judge the recovery of the planetary signals,
		we set a threshold with the SDE of 7 based on the past works in the literature.}
		Empirical thresholds are set between 6 and 10 \citep[e.g.,][]{Dressing2015, Wells2018}.



	\begin{figure*}[t]
		\centering
		\includegraphics[width=16cm]{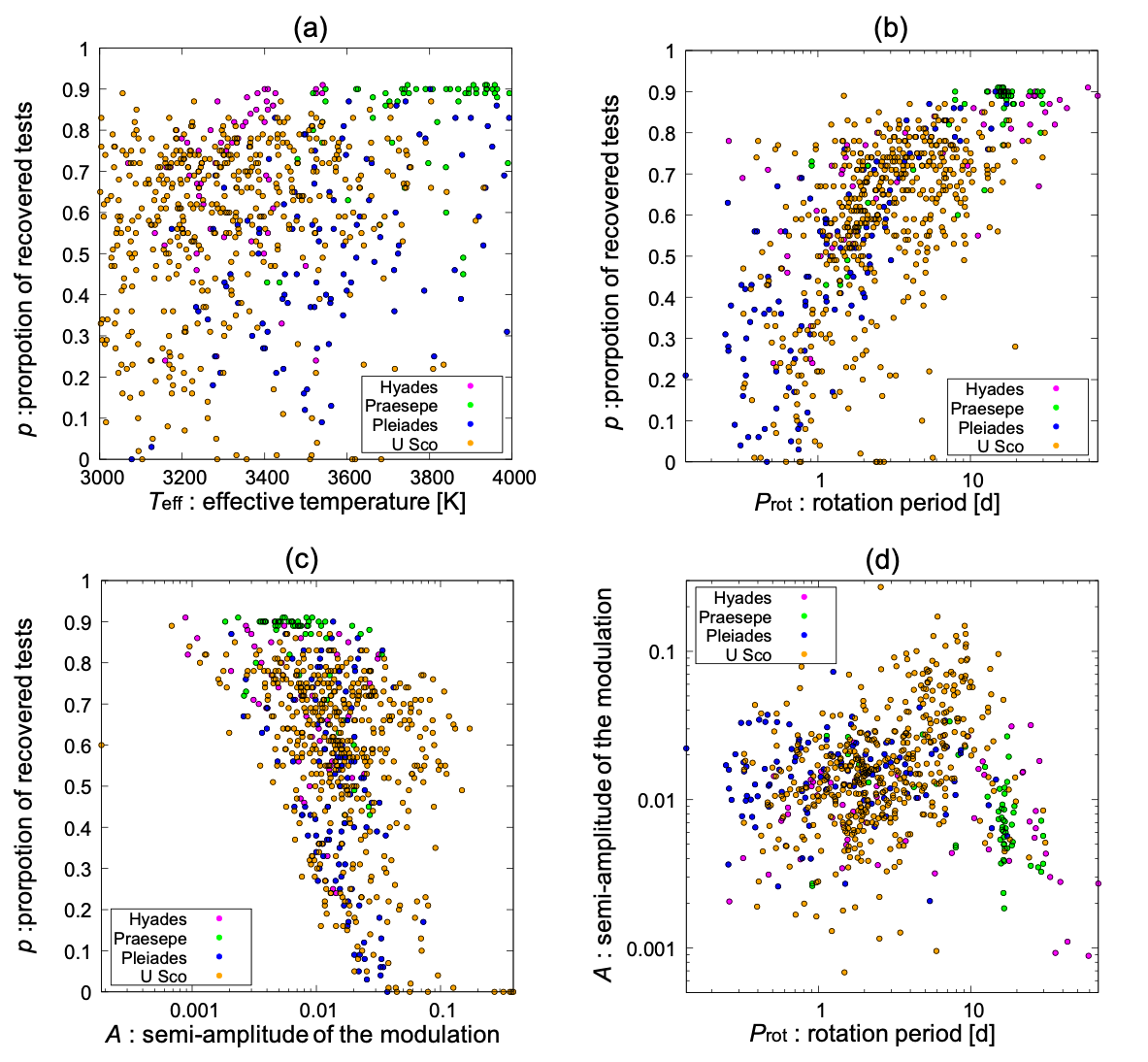}
		\caption{Scatter plots colored by cluster.
		Panels (a), (b), and (c) plot the recovery rate $p$,
		{ the relative number of recovered tests to all tests using the 10 $\times$ 10 planetary parameters,}
		vs. effective temperature, rotation period, and semi-amplitude of the photometric variation
		of the host stars, respectively.
		(d) plots the semi-amplitude of the variation vs.  the rotation period.}
		\label{pic:scatter}
	\end{figure*}

	\begin{figure}[t]
		\centering
		\includegraphics[width=9cm]{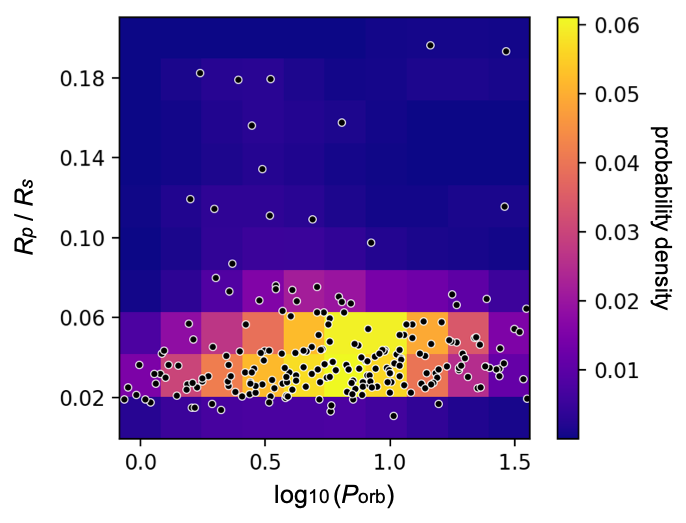}
		\caption{Probability density function (PDF) for planets around stars
		whose $T_{\rm eff}$ are 3000 - 4000 K discovered by the transit method.
		The black dots are orbital property of the planets.
		The probability density is calculated by the Gaussian kernel density estimation.}
		\label{pic:current_pdf}
	\end{figure}
	\begin{figure}[t]
		\centering
		\includegraphics[width=8cm]{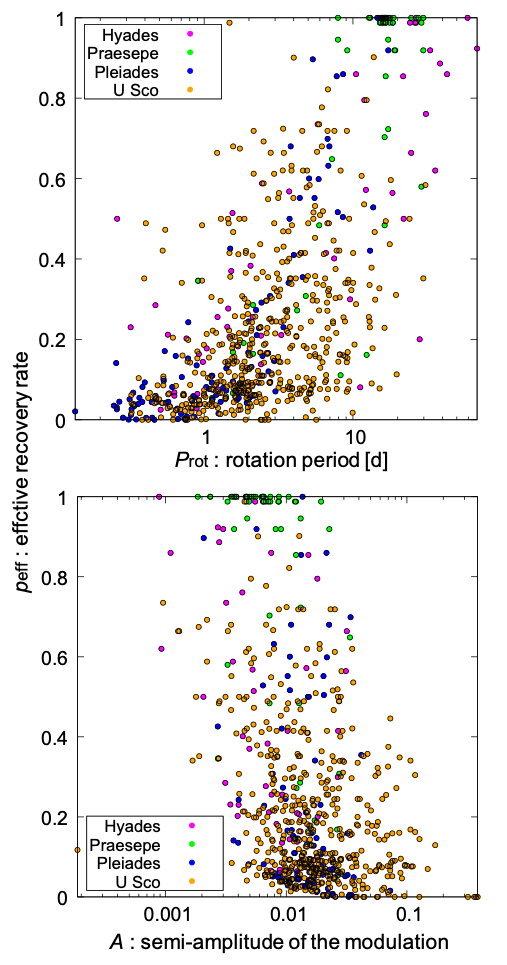}
		\caption{Scatter plots of $p_{\rm eff}$, effective recovery rate based on
		the current planet distribution, colored by cluster.
		The upper and lower panels are the detectability assuming the PDF in Figure \ref{pic:current_pdf}
		to the rotation period and semi-amplitude of the flux modulation, respectively.}
		\label{pic:scatter_current}
	\end{figure}


	%

\section{Results}\label{sec:res}
	\subsection{SDE Maps of Typical Targets}

	Through the injection and recovery tests described in Section \ref{sec:method},
	we visualized the detectability of the planetary signals as SDE maps of the $10 \times 10$ orbital parameters in Figure \ref{pic:cmap}.
	The vertical and horizontal dotted lines represent the rotation
	period and the square root of the flux variation semi-amplitude, respectively;
	the color bar corresponds to the SDE in the TLS test of each mock light curve;
	the white lines are the boundary of SDE = 7 as the detection threshold.
	The targets in the upper and lower rows show large and small
	amplitude variations, respectively.
	The rotation periods of the leftmost, center left, center right, and rightmost panels were approximately
	10, 3, 1, and 0.3 days, respectively.
	{ In the top of each panel, we put the photometric variation amplitude $A$,
	rotation period $P_{\rm rot}$, and mean of the standard deviation
	in 0.3 d - sized bins $\sigma_{\rm 0.3 d}$ as a rough standard of the photometric precision.}
	\begin{figure*}[t]
		\centering
		\includegraphics[width=18cm]{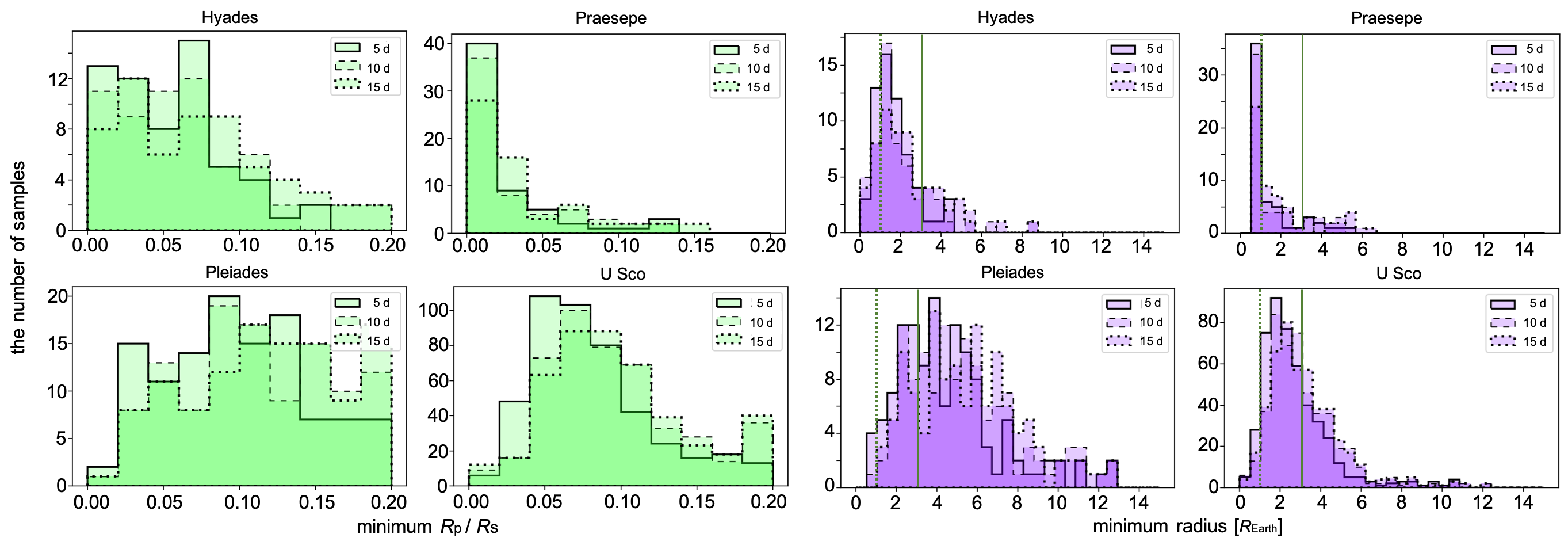}
		\caption{Histograms showing the number distributions of minimum
		detectable radius ratio $R_{\rm p}/ R_{\rm s}$ (left)
		and planetary radius $R_{\rm p} [R_{\oplus}]$ (right)
		with SDE = 7 for a given orbital period in each cluster.
		In the right panels, dotted and dashed vertical lines denote the Earth and Neptune radii, respectively.
		The solid, dashed, and dotted lines denote different orbital periods of the injected planets (5, 10, and 15 d, respectively).
		}
		\label{pic:res_hist}
	\end{figure*}

	As shown in the upper panels, the detectability of the planetary candidates 
	{decreases with increasing rotational velocity.}
	For the slowest rotator (EPIC 211000411),
	a radius ratio $R_{\rm p}/R_{\rm s}$ of 0.02, corresponding to a transit depth of $0.04\%$,
	is detectable.
	{In this case, the detectability appears to be dominated
	by photometric precision, whose influence is discussed in \citet{Nardiello2021},
	rather than by stellar activity.}
	In contrast, a planet around the rapid rotator with $P_{\rm rot}$ of $\leq$ 1 d (EPIC 210947895)
	is difficult to detect
	even at $R_{\rm p}/R_{\rm s}$ of 0.1 (= $1 \%$ transit depth).
	In the lower panels, the detection limit for rapid rotators lies
	around $R_{\rm p}/R_{\rm s}$ = 0.05 ($\approx 0.3 \%$ transit depth).
	These boundaries { vary corresponding to the amplitudes of variations }
	caused by stellar activity { compared to the upper panels}.

	As a benchmark (black star in Figure \ref{pic:cmap}), we selected
	a known planetary system orbiting a M-dwarf,
	K2-25b \citep[EPIC 210490365b; ][]{Mann2016},
	whose host star property is similar to EPIC 211120664.
	The relative planetary radius, orbital period, stellar rotation period,
	and amplitude of variation of the K2-25 system
	are approximately 0.1, 3.5 d, 1.9 d, and $1 \%$, respectively.
	Although K2-25b  orbits a very rapidly rotating M-dwarf,
	its rotational period and semi-amplitude of the variation are set significantly above the detection limit.
	Our results also indicate that if other Earth-sized planets exist around K2-25,
	they will be obscured by stellar activity.

	Stellar activity appears to affect the detectability of planetary candidates
	especially when the stellar rotation periods are shorter than approximately three days.
	Pleiades and Upper Scorpius include
	a large number of such rapidly rotating stars \citep{Rebull2016, Rebull2018}.
	In Hyades and Praesepe, there are large scatters of rotation periods in low-mass regions,
	and some targets (possibly binaries) rotate with periods shorter than three days \citep{Douglas2019}.
	{ Even when the variation amplitudes of the host stars are small (lower row of Figure \ref{pic:cmap}),
	Earth-sized planets tend to be missed around the rapidly rotators with $P_{\rm rot} \leq 1$ d}.
	TRAPPIST-1f, a habitable Earth-sized planet
	with a relative radius $R_{\rm p}/R_{\rm s}$ and
	orbital period of approximately 0.08 and 9 days, respectively,
	would not be detected in the TLS analysis,
	if the planet exists around active stars like
	EPIC 210892390, EPIC 210947895, and EPIC 211098117.
	Nonetheless, we are aware of sophisticated methods or optimizations \citep[e.g.,][]{Rizzuto2017}
	that enhance the detectability of small planets around active stars.


	\begin{figure}[h]
		\centering
		\includegraphics[width=7.7cm]{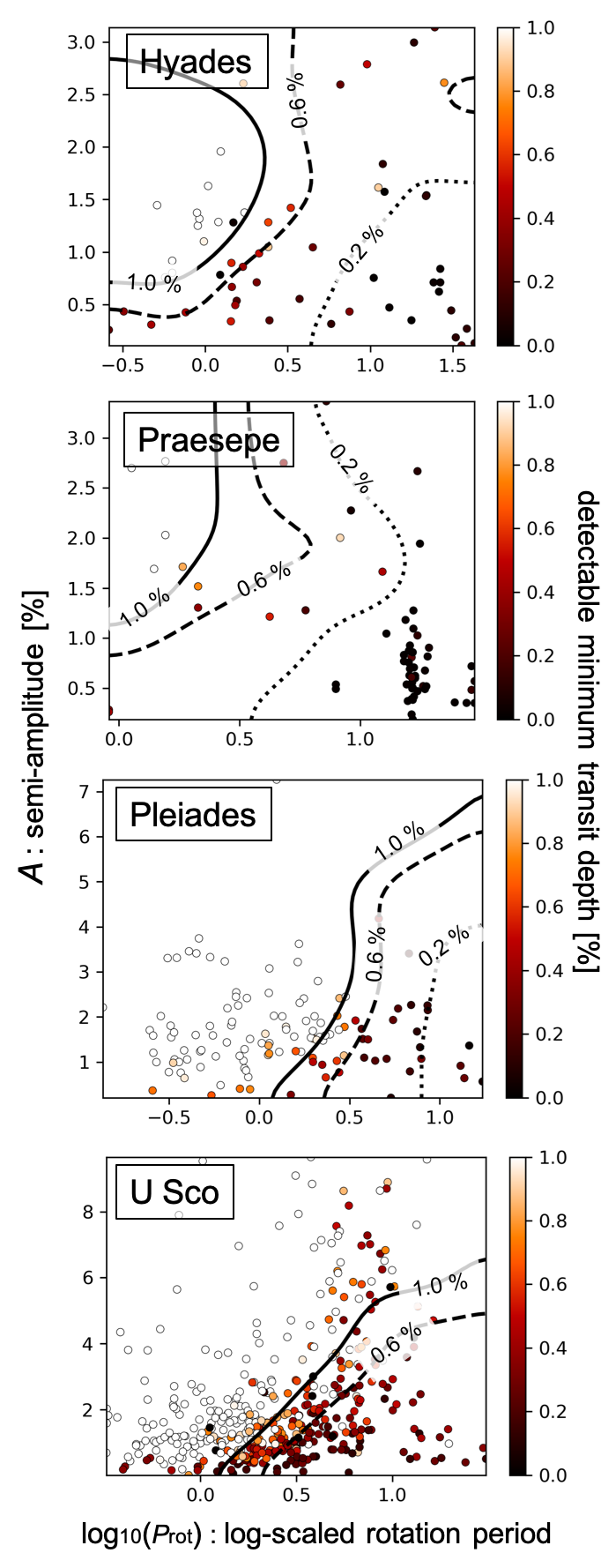}
		\caption{
		Scatter plot of the rotation period versus semi-amplitude of photometric variation
		for each cluster. The color bar corresponds to the minimum detectable transit depth [$(R_{\rm p}/R_{\rm s})^2\%$]
		when the orbital period is 10 d.
		{The contour lines represent borders of 1.0\%, 0.6\%, and 0.2\%
		detectable minimum transit depth.}
		e.g., The targets existing above the border of 1.0\% could not be detected
		when the transit depth is shallower than 1.0\%.}
		\label{pic:res_vis_P10}
	\end{figure}
	\subsection{Trends in the Simulations for the Whole Sample}\label{sec:totaltrend}
	{We show { the qualitative dependence of the test results on }
	rotation period, and amplitude of the flux modulation to confirm correlations among the clusters.}
	We calculate the proportion of recovered cases $p$ as a guide for interpretations.
	Here, $p$ defines the relative number of recovered tests
	to all tests using the 10 $\times$ 10 planetary parameters for each sample;
	$p = 0$ and $p = 1$ indicate that no planet is recovered and
	that all planets are recovered, respectively.
	of Figure \ref{pic:scatter}-(a) displays the recovery-rate distribution for each cluster
	with respect to the effective temperature of the host star.
	The $T_{\rm eff}$ biases are likely introduced by distances to the cluster,
	because the samples were selected based on their apparent magnitudes.
	The samples in Pleiades and Upper Scorpius are more scattered along both the effective temperature
	and $p$ axes than those in Hyades and Praesepe.
	This difference may be caused by the different number of samples in each cluster.
	Despite these selection biases,
	the $p$ and the $T_{\rm eff}$ are apparently uncorrelated.

	Figure \ref{pic:scatter}-(b) and (c)
	display $p$ with respect to the rotation period $P_{\rm rot}$
	and the semi-amplitudes of the light curve $A$, respectively.
	Unlike $T_{\rm eff}$,  both $P_{\rm rot }$ and $A$ appear to influence the detectability $p$
	with similar trends for each cluster.
	As shown in the scatter plot of $P_{\rm rot}$ versus $A$ (Figure \ref{pic:scatter} (d)),
	these two parameters are not clearly correlated despite their biased distributions in each cluster.
	Therefore, the trends displayed in panels (b) and (c) are independent.
	{ As a reference, the correlation coefficients
	are 0.49, -0.33, and -0.03 for panels (b), (c), and (d), respectively.}
	{ Finally, we conclude that the detectability of the transit signals depends on the rotation period
	and amplitude of the flux modulation, respectively, rather than the effective temperature
	or cluster where the stars belong.
	{ Note that we do not discuss the detectability assuming a realistic exoplanet distribution,
	but a statistical trend in our simulation here.}
	}

	\subsection{Detectability Assuming the Observed Planetary Distribution}\label{sec:detectability_current}
	{
	In this subsection, we derive { an effective recovery rate $p_{\rm eff}$, 
	a proportion of detectable planets when one transiting planet exists around each sample based on
	the current planet distribution.}
	We downloaded a list of the orbital parameters of confirmed planets
	which were detected by the transit method
	around stars with $T_{\rm eff} $ of 3000 - 4000 K
	from \citet{confirmed} \footnote{https://exoplanetarchive.ipac.caltech.edu/}.
	After removing the young planets listed in Table \ref{tab:planets} from the list,
	we calculated a probability density function (PDF) for 249 confirmed planets
	by the Gaussian kernel density estimation
	which is shown in Figure \ref{pic:current_pdf}.
	{We derive the $p_{\rm eff}$ based on the observed planet distribution
	by multiplying the PDF by the detectability of the stellar activity for each pixel
	as $\delta_{\rm SDE} = \{\rm 1 ~(SDE \geq 7),~ 0 ~(SDE < 7)\} $
	and integrating over the orbital parameter region. }
	Because the PDF is likely affected by the photometric precision,
	we avoid using a region where the radius ratio is less than 0.02 for this calculation.

	{ We show the derived { effective recovery rate $p_{\rm eff}$}
	to the rotation period and semi-amplitude of the flux modulation
	in the upper and lower panels of Figure \ref{pic:scatter_current}, respectively.
	Compared with Figure \ref{pic:scatter} (b) and (c),
	samples seem to be biased toward the region where the $p_{\rm eff}$ is less than 0.4.
	This declination of $p_{\rm eff}$ is due to that the planetary distribution
	is dense around $R_{\rm p}/R_{\rm s} = $ 0.02 - 0.06 and $P_{\rm rot} = $ 5 - 10 d.}
	{ The medians and ranges of 68 \% distribution between stars of $p_{\rm eff}$ value
	are about $0.33 _{-0.25}^{+0.58}$, $0.99_{-0.72}^{+0.01}$,
	$0.08_{-0.05}^{+0.42}$, and $0.16_{-0.11}^{+0.29}$ }
	for Hyades, Praesepe, Pleiades, and Upper Scorpius, respectively.
	Note that these calculations are performed on assumption that
	the young planetary distribution is the same as the older counterpart.
	}

	\subsection{Systematics of Each Cluster}
	Section \ref{sec:totaltrend} presented the statistical properties of the whole filtered samples.
	We now focus on the results for each cluster.
	Figure \ref{pic:res_hist} shows histograms of the numbers of samples
	with respect to the minimum detectable radius ratio and converted radius.

	Although the detectability decreases with increasing orbital period, the degradation is slight.
	For most of the samples in Hyades, Praesepe, and Upper Scorpius,
	the minimum $R_{\rm p}/R_{\rm s}$ detectable is less than 0.10 (left panels of Figure \ref{pic:res_hist}).
	The high detectability in Praesepe is attributable to the selection bias.
	As shown in Figure \ref{pic:hist},
	we selected relatively bright targets and stars with high effective temperatures,
	for which the signal modulations are stable \citep{Rebull2018}.
	Although a similar selection bias exists in Pleiades targets,
	the distribution of detection limits in Pleiades is widely scattered,
	indicating that most of the Pleiades samples are more active than those of Praesepe.

	\begin{figure*}[ht]
			\centering
			\includegraphics[width=18.5cm]{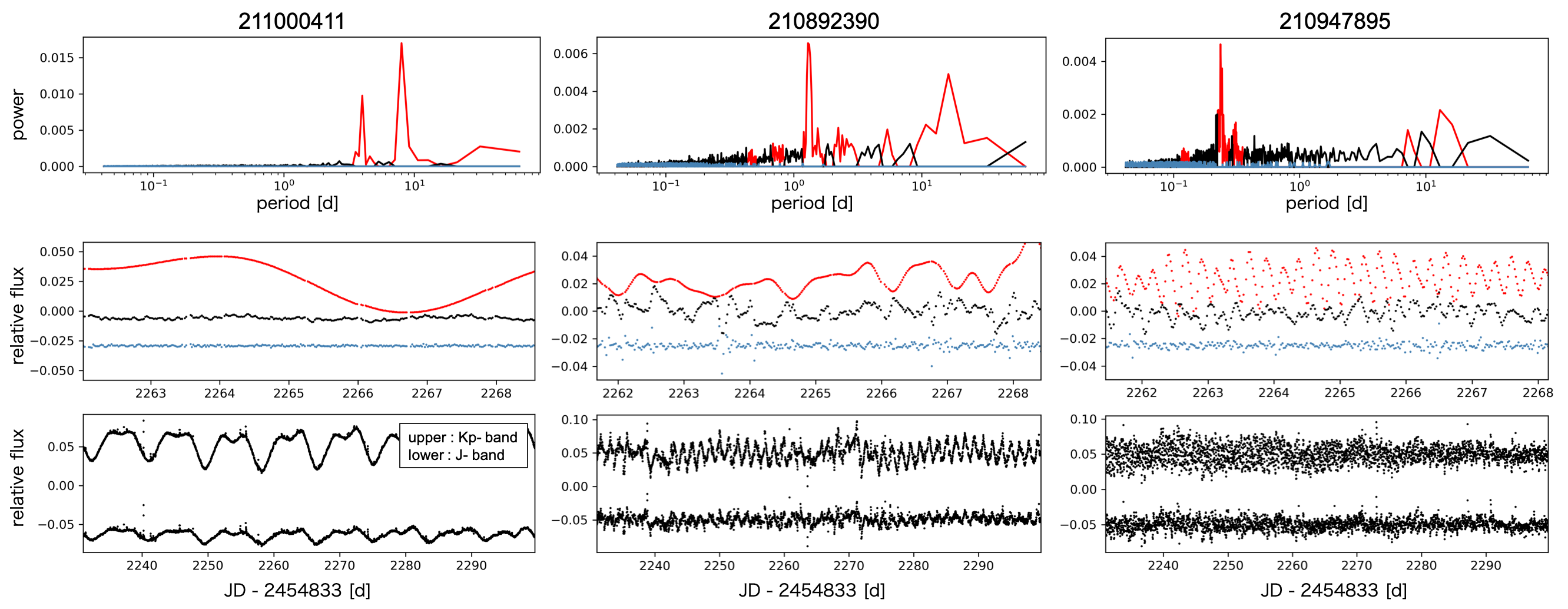}
			\caption{Explanations of the analysis in subsection \ref{sec:nir_rep} for three typical targets.
			For each target, the top panel plot the periodogram derived by FFT.
			The red, blue, and black data represent
			sinusoidal modulation detected by LS, white noise, and other correlated noise,
			respectively.
			In the middle panel, the range is narrowed to approximately five days
			to clarify the details of light curves reproduced by inverse FFT.
			The color relation is same to the top panel.
			The bottom panels plot the original light curves in the $Kp$-band (upper)
			and reproduced light curves assuming the $J$-band observation (lower).}
			\label{pic:FFT}
	\end{figure*}
	\begin{figure}[h]
			\centering
			\includegraphics[width=9cm]{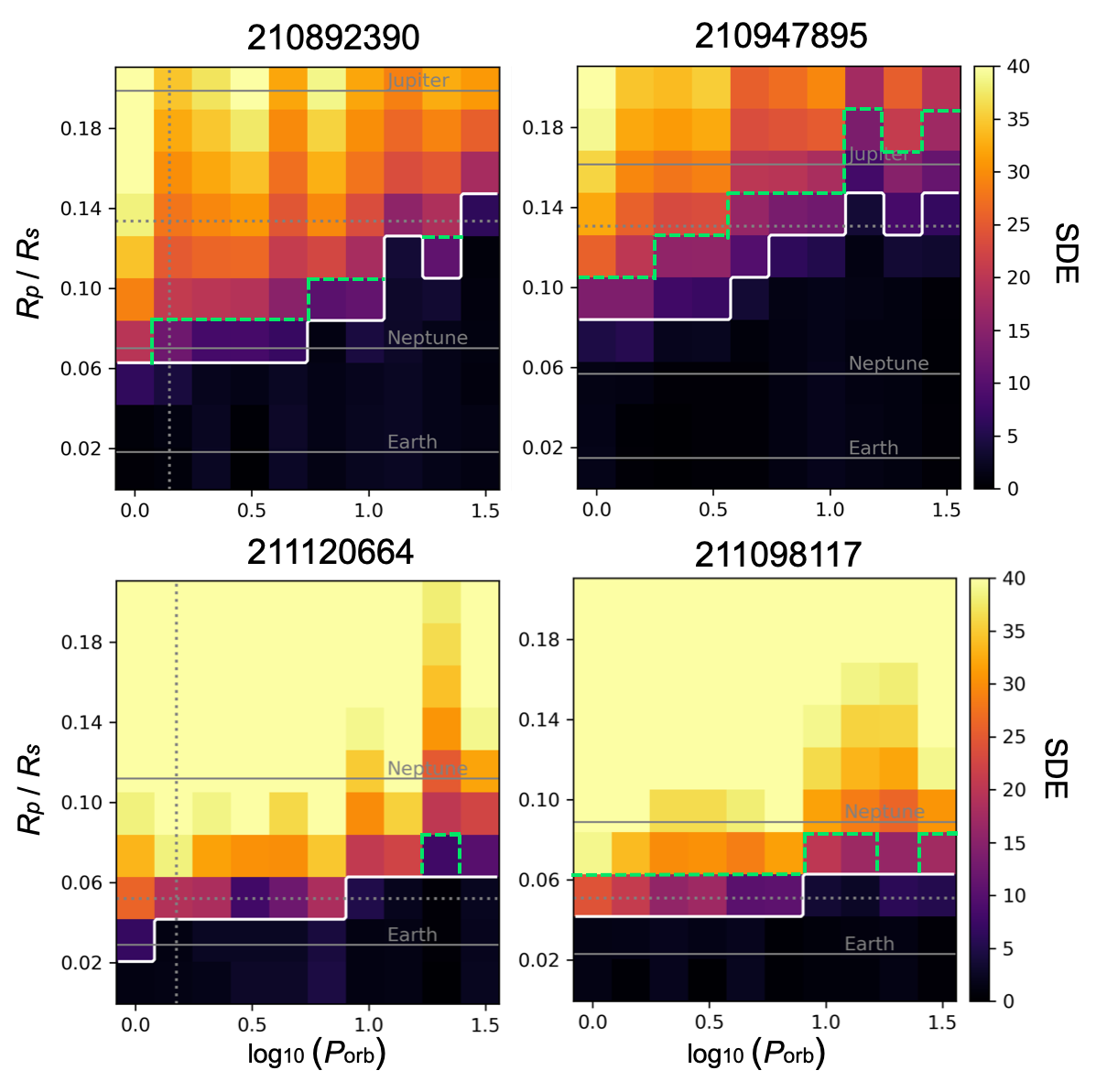}
			\caption{Same heat maps with the right side of Figure \ref{pic:cmap}, but in the $J$- band.
				The dashed green lines represent the border in the $Kp$- band of Figure \ref{pic:cmap}
				for comparison.}
			\label{pic:cmap2}
	\end{figure}

	To assist the astrophysical interpretation,
	we present the same histograms against the planetary radius
	in the right panels of Figure \ref{pic:res_hist}.
	{ For the histograms with the orbital period of 10 d,
	we derived { the median planet radius} and the fraction of stars { for} which Earth-size
	and Neptune-size planets can be detected ($f_{\rm Earth}$, $f_{\rm Neptune}$)
	for each cluster in Table \ref{tab:res_hist}.}
	Neptune-sized planets are detected around most targets in Hyades and Praesepe,
	and a reasonable {rate} of samples passed the threshold of Earth-sized planets.
	In Upper Scorpius samples, most injected planets larger than Neptune were also detected.
	Conversely, planet detection among the Pleiades members was difficult
	and very few Earth-sized planets are detectable.
	{ Even for Neptune-sized planets, the passed fraction is only 20 \%.}
	{ Note that we may possibly underestimate the planetary distribution
	due to the inflation of stellar radius in the early stage
	and the detection in Pleiades and Upper Scorpius may be more difficult than we estimated.}

	In addition, we derived p-values by performing the Kolmogorov-Smirnov test
	for the histograms with the orbital period of 10 d to the Hyades, Praesepe, and Pleiades
	($\rm P_{KS, Hya}$, $\rm P_{KS, Pra}$, and $\rm P_{KS, Ple}$)
	and listed them in Table \ref{tab:res_hist}.
	{ The distributions are different each other
	according to the small P-values ($< 0.05$). }

	Figure \ref{pic:res_vis_P10} clarifies the relation
	between the detectability and light curve modulation.
	In this figure, the white circles represent the targets
	whose transit depths of less than 1\% cannot be detected
	if planets exist around them.
	The detection limits have a trend from the left bottom to the right top in the plots of all clusters.
	{We also show the contours which represent borders of 1.0 \%, 0.6 \%, and 0.2 \%
	of the minimum detectable transit depths.
	The contours were estimated with Gaussian filtering with 1 $\sigma$ of 5 pix
	after the nearest 2D interpolation of the parameter space with 50 $\times$ 50 grids
	for the scattered data points.}
	This systematic behavior can explain the distributions of detectable $R_{\rm p}/R_{\rm s}$ for each cluster
	in Figure \ref{pic:res_hist},
	indicating that the detectability in Pleiades simply depends on the light curve feature.
	In general, cool stars rotate more rapidly than hot stars in their young phase.
	We selected relatively bright Pleiades members (see Figure \ref{pic:hist}),
	and other Pleiades members may be cooler and more rapid rotators than our targets.
	{ Thus, it will be more difficult to detect planetary transit signals for all the Pleiades members
	than we estimate in this study.}
	Consistently with this assumption, no secure detection of transiting planets
	has been reported for Pleiades members \citep{Gaidos2017}.

	\section{Expectation in NIR Observations}\label{sec:nir}
	\begin{figure*}[t]
			\centering
			\includegraphics[width=18cm]{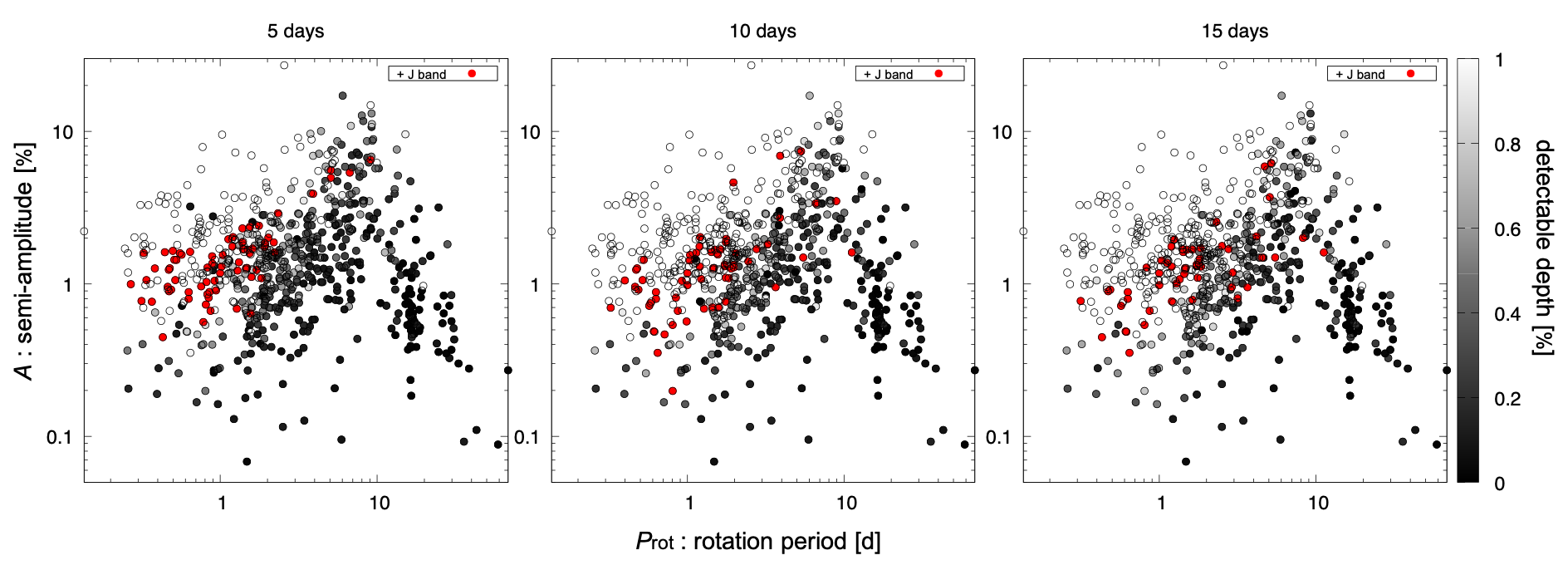}
			\caption{Scatter plots of the semi-amplitude of photometric
			variation and detectable depth versus rotation period.
			The color bar presents the minimum detectable transit depth in the $Kp$- band.
			The red points are the detectable targets at a threshold of $1 \%$
			in the $J$- band, but not in the $Kp$- band.
			In the left, middle, and right panels, the planetary orbital periods of samples are
			5, 10, and 15 days, respectively.}
			\label{pic:res_1percent}
	\end{figure*}

	\begin{table}
		\center
		\caption{Statistical values for the histogram in the right of Figure \ref{pic:res_hist}.}
		\begin{tabular}{ccccc}
			\hline \label{tab:res_hist}
			&Hyades & Praesepe & Pleiades & U Sco \\
			\hline
			${\rm P_{KS, Hya}}$ & - & $4.9 \times 10^{-5}$ & $3.0 \times 10^{-15}$ & $3.7 \times10^{-8}$ \\
			${\rm P_{KS, Pra}}$ & - & - &$4.4\times10^{-16}$ &$3.0\times10^{-17}$\\
			${\rm P_{KS, Ple}}$ & - & - &- &$2.1\times10^{-15}$ \\
			median [$R_\oplus$]&1.6&0.9&4.9&2.8\\
			$f_{\rm Earth}$ &0.20&0.60&0.01&0.04\\
			$f_{\rm Neptune}$ &0.80&0.84&0.20&0.56\\
			\hline
		\end{tabular}
	\end{table}

	\subsection{Reproduction of the NIR Photometry}\label{sec:nir_rep}
		As mentioned in Section \ref{sec:intro}, the light curve variations caused by stellar activity are
		wavelength dependent.
		{ Some studies suggested a trend that
		the amplitude of the sinusoidal flux modulation is  
		lower at NIR than at visual wavelengths by about 50 - 80 \% for M-dwarfs
		\citep[e.g., ][]{Frasca2009, Morris2018, Miyakawa2021b}.}
		If this is the case for most active low-mass stars,
		then planetary detection around young targets should be enhanced
		in future transit search with NIR instruments.
		When evaluating the wavelength dependency of the detected transit signals
		and the efficiency of the NIR survey,
		we performed the test assuming observations in the NIR.

		We first applied the FFT
		to the light curves observed with the $Kp$ - band in the {\it K2} mission.
		To detect the frequencies of
		the sinusoidal modulation (i.e., stellar jitter),
		we also computed the LS periodogram.
		The amplitudes of the modulations corresponding to significant peaks in the LS periodogram
		(within $10 \%$ uncertainty of the peaks detected in the LS
		and with false alarm probabilities below $1\%$)
		were adjusted to the $J$ - band observations.
		Here, the amplitude ratio of the $J$ to $Kp$ - bands,
		was assumed as $34.8\%$,
		the observational values derived as the mean of the four active M-dwarfs belonging to the Hyades
		in \citet{Miyakawa2021b}.

		The white noise was determined as signals with powers lower than
		$F_{\rm med} + 3 \sigma$ in the FFT spectrum,
		where $F_{\rm med}$ and $\sigma$ are the median and standard deviation, respectively of
		the signals with frequencies exceeding $15 ~{\rm d}^{-1}$.
		In the $J$ - band, we re-scaled the white-noise scatter in the $Kp$ - band photometry
		based on the photon counts derived from the spectral energy distribution (SED),
		applying the PHOENIX spectra of \citet{Allard2013} as the SED model.
		The photometric flux for a given effective temperature of the host star
		was calculated as described in \citet{Fukugita1995}.
		{ Note that we assume the same telescope profile with  the {\it Kepler}
		because we focus on the wavelength dependency of the signal.}
		After reproducing the NIR light curves through the above procedures,
		we applied the tests described in subsection \ref{sec:injection} and \ref{sec:recovery}.

		Figure \ref{pic:FFT} demonstrates the analysis on three typical targets;
		EPIC 21100411, EPIC 210892390, and EPIC 210947895
		with rotational periods of $\sim10$ d, $\sim1$ d, and $\sim0.3$ d, respectively.
		In the FFT results of each target (upper panels),
		the red, blue, and black peaks were detected as astrophysical signals in the LS periodogram,
		white noise, and other correlated noise, respectively
		and the middle panels displays the corresponding light curves at
		the frequencies discriminated in the FFT of each target.
		The correlated noise for the rapid rotators (EPIC 210892390 and EPIC 210947895)
		is significantly larger than that of EPIC 21100411,
		possibly because
		the correlated noise is not easily distinguished from rotational modulation
		in the periodogram analysis.
		Consequently, the rotational modulations were underestimated,
		but this uncertainty did not affect our estimations of the lower limits
		of improvement in the mock $J$ - band observation.
		The multi-periodicities in EPIC 210892390 and EPIC 210947895
		might be explained by differential rotation and/or binarity.
		In either case, the wavelength dependency of the rotational modulation
		will probably not vary significantly from the basic assumption.
		The bottom panels of Figure \ref{pic:FFT} show the original light curves and the mock light curves
		assuming $J$-band observations.


	\subsection{Improvement in NIR}

	The SDE maps of the rapid rotators EPIC 210892390, EPIC 210947895, EPIC 211120664,
	and EPIC 211098117 shown in Figure \ref{pic:cmap}
	are replotted assuming NIR observations in Figure \ref{pic:cmap2}.
	Also plotted in Figure \ref{pic:cmap2} are the borders in $Kp$- band.
	Although the detectability was not apparently improved around slowly rotating ($P_{\rm rot} \sim10$) -
	and less active ($A \sim0.1\%$) -stars,
	the detectable relative radius $R_{\rm p}/R_{\rm s}$ was improved by at most 0.04 (corresponding to two grid widths)
	for the very active targets.

	Figure \ref{pic:res_1percent} plots the semi-amplitude of the photometric variation and
	minimum detectable transit depth as functions of
	stellar rotation period for all targets derived in the injection and recovery tests.
	The left, middle, and right panels correspond to
	planetary orbital periods of 5, 10, and 15 days, respectively.
	Some targets (highlighted in red) were detectable at a threshold of $1\%$ in the  $J$- band
	but not in the $Kp$- band.
	The limits in both $Kp$ and $J$- bands deteriorate with increasing orbital period,
	{ because the number of transit signals decreases in a given observation span}.
	{
	Targets that were detectable only in the $J$-band (red points)
	appear to be distributed in the upper left of those that
	were detectable in the $Kp$ band (gray - black points),
	regardless of orbital periods.
	Such targets are likely members of Pleiades and Upper Scorpius,
	where active targets dominate (see Figure \ref{pic:scatter}-(d)).}

\begin{figure*}[ht]
			\centering
			\includegraphics[width=18cm]{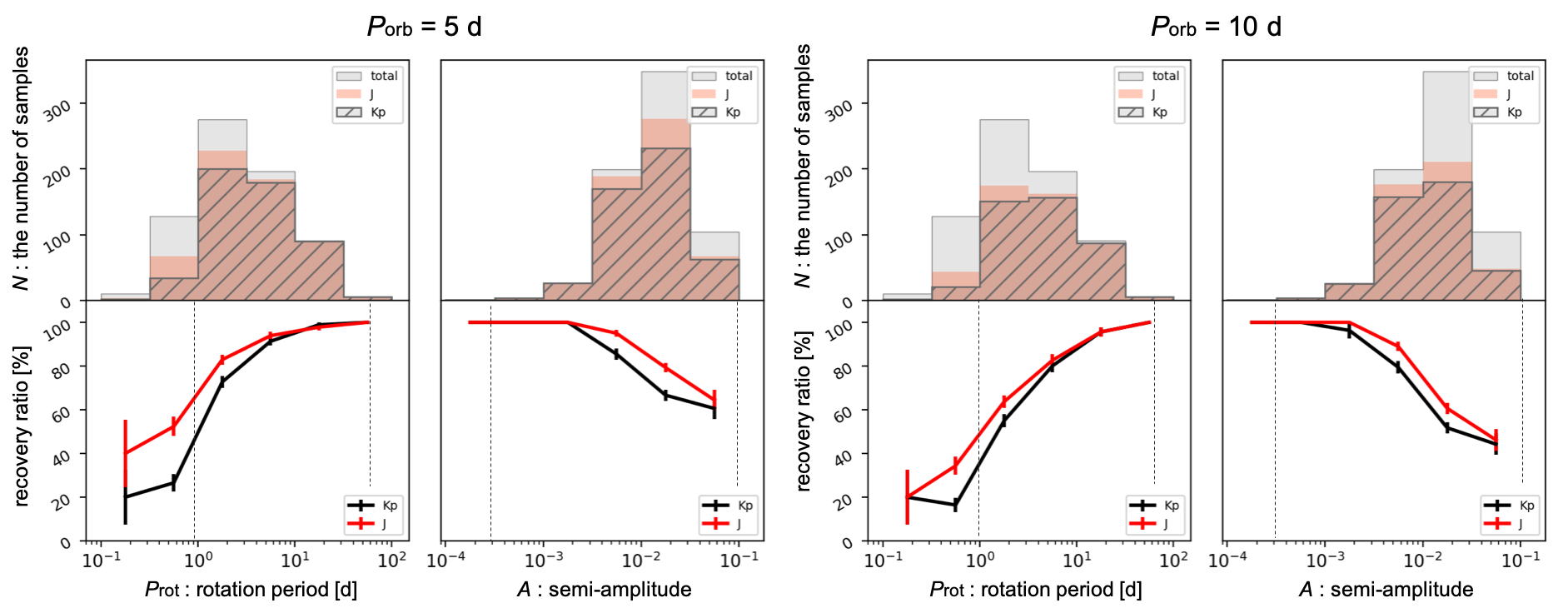}
			\caption{Top panels: histogram plot of Figure \ref{pic:res_1percent}
			with respect to the rotation period and semi-amplitude of the photometric variation.
			{All targets and targets that are detectable in the $J$- and $Kp$- bands with threshold of 1 \%
			are shown with gray, red and black diagonals, respectively.}
			Bottom panels: the rates of detectable number to the total number of samples
			in the $J$- and $Kp$- bands (red and black lines, respectively) corresponding to the upper panels.
			The vertical dashed lines represent parameter ranges shown in Figure \ref{pic:current}.}
			\label{pic:number_of_samples}
	\end{figure*}

	To evaluate the detectability, we plot the accumulated distribution in Figure \ref{pic:res_1percent}
	as functions of $P_{\rm rot}$  and $A$ (Figure \ref{pic:number_of_samples}, upper panels)
	and the ratio of detectable samples to all samples in the $J$- and $Kp$- bands
	(Figure \ref{pic:number_of_samples}, lower panels)
	for planetary orbital periods of five days (left) and 10 days (right).
	Regardless of orbital period, the differences between the $Kp$- and $J$- bands
	is small at rotation periods around ten days,
	but the recovery rates were almost doubled in the rapidly rotating region ($P_{\rm rot}$ $\sim 1 $ day).
	Meanwhile, the detectability against the amplitude $A$ was improved by at most 10 \% in the $J$- band
	relative to the $Kp$- band.
	\begin{table*}
		\centering
		\caption{ Summary of the discussion in Section \ref{sec:clusters}.}
		 \label{tab:clusters}
		\begin{tabular}{ccccc}
		\hline \hline
				& Hyades & Praesepe & Pleiades & Upper Scorpius \\ \hline
			\multicolumn{5}{c}{basic information} \\
			typical age [Myr] & 650 - 700 & 650 - 700 & 112 -125 & 8 - 11 \\
			typical {\it Kp}- magnitude of M-dwarfs & $13.84_{-0.92}^{+0.92}$ & $16.92_{-1.75}^{+1.46}$
				& $15.69_{-1.06}^{+0.87}$& $13.94_{-0.90}^{+0.94}$ \\
			typical $P_{\rm rot}$ [d] of M-dwarfs & $2.45_{-1.69}^{+22.19}$ & $2.06_{-1.47}^{+16.49}$
				& $0.63_{-0.33}^{+1.19}$ & $2.32_{-1.52}^{+5.09}$\\
			(i) the number of M-dwarfs  & 164 & 794 & 759 & 969\\
			(ii) the number of planets with  $P_{\rm rot}$ $\leq$ 10 d of M-dwarfs& 1& 3& 0& 1\\ \hline
			\multicolumn{5}{c}{qualitative estimation in Section \ref{sec:clusters}.}\\
			(iii) planet frequency around M-dwarf by (ii)/(i) & 0.006& 0.004& 0.000& 0.001\\
			(iv) detection effciency with {\it Kp}-magnitude & 0.23& 0.18& 0.21& 0.23\\
			(v) detection efficiency with $P_{\rm rot}$ in {\it Kp} & $0.72\pm0.03$& $0.72\pm0.03$
				& $0.27\pm0.04$& $0.72\pm0.03$\\
			normalized planet frequency by (iii)/((iv)$\times$(v))& $0.035\pm0.008$& $0.031\pm0.009$
				& $0.000$&$0.006\pm0.001$ \\ \hline \hline
		\end{tabular}
	\end{table*}

	\section{Discussions}\label{sec:discussion}
	\subsection{Implication to the Clusters }\label{sec:clusters}
	We { summarize} the planetary frequencies in the clusters, Hyades, Praesepe, Pleiades, and Upper Scorpius
	considering the detection efficiency with respect to the rotation period
	in Figure \ref{pic:number_of_samples}.
	Below, the values are shown in the order of Hyades, Praesepe, Pleiades, and Upper Scorpius.
	{ Firstly, }as in Table \ref{tab:planets}, currently 1, 3, 0, and 1 planetary systems were
	reported around M-dwarfs with effective temperature of 3000 - 4000 K and
	orbital period of $\leq$ 10 d.
	We put asterisks to highlight these planetary systems in Table \ref{tab:planets}.
	The numbers of M-dwarf members in the clusters reported in the literature \citep{Rebull2016,
	Douglas2017, Rebull2018, Douglas2019}
	are 164, 794, 759, and 969.
	Although the number of detected planets is statistically poor,
	the frequency of the detected planets around M-dwarfs with close-in orbit
	are simply 0.006, 0.004, 0.000, and 0.001.
	{ Next, we consider the planet detectability with stellar brightness for each cluster,
	because we only discussed that with stellar activity in a given magnitude range in the previous sections.}
	{ The medians and 63 \% errors of apparent magnitude of apparent
	{\it Kp} magnitude of M-dwarfs in the clusters are
	$13.84 _{-0.92}^{+0.92}$, $16.92_{-1.75}^{+1.46}$, $15.69_{-1.06}^{+0.87}$, and $13.94_{-0.90}^{+0.94}$.}
	Figure 6 in \citet{Nardiello2021} proposed that
	the { nomalized detection efficiencies} thorough injection-recovery tests using {\it TESS} light curves
	{ for different apparent magnitudes} are approximately 0.23, 0.18, 0.21, and 0.23
	for  ${\rm mag}_{T}$ of 12 - 13, 15 - 16, 14 - 15, and 12 - 13, respectively,
	for the systems with planetary radius of 3.9 - 11.2 $R_{\oplus}$ and orbital period of 2 - 10 d.
	Here we assume that the magnitude difference between $Kp$ - and $T$ - bands is about 1
	when the stellar mass is around 0.4 $M_{\odot}$
	based on the MESA Isochrones and Stellar Tracks \citep[MIST; ][]{Dotter2016, Choi2016}.
	{ Then, we derive detection efficiency with stellar rotation period
	referring the results in Figure \ref{pic:number_of_samples}.}
	Because the typical rotation periods { for the four} clusters are 2.5 d, 2.1 d, 0.6 d, and 2.3 d,
	the qualitative detection efficiencies are
	{$0.72 \pm 0.03$, $0.72 \pm 0.03$, $0.27 \pm 0.04$, and $0.72 \pm 0.03$}, assuming
	the orbital period of 5 d in Figure \ref{pic:number_of_samples}, respectively.
	Thus, the detectabilities considering both the apparent magnitude and rotation period
	{ are $0.17 \pm 0.04$, $0.13 \pm 0.04$, $0.06 \pm 0.02$, and $0.17 \pm 0.04$
	by multiplying these values,
	where we { assume nominal error value of 0.05 for the detection efficiencies with apparent magnitude.}}
	These detectabilties seem to explain the current no planet detection in Pleiades \citep{Gaidos2017}.
	{ Finally}, normalized planet frequencies by these detection efficiencies
	{ are approximately $0.035 \pm 0.008$, $0.031 \pm 0.009$, $0.000$, and $0.006 \pm 0.001$.}
	This result indicates that potential planets in Upper Scorpius are likely lacking compared
	with Hyades and Praesepe due to evolution events.
	For example, the planet migration might occur after Upper Scorpius age around M-dwarfs.
	{ Note that we do not consider uncertainties in the planetary frequencies and
	may underestimate errors in this assessment because we use a given value as the detectability
	while the stellar properties have a large scatter for each cluster.
	{ We summarize these qualitative discussion in Table \ref{tab:clusters}.}

	Besides, we derived 
	{ the typical values of the effective recovery rate $p_{\rm eff}$,
	which can be also interpreted as the detectability { for different} stellar activity level
	assuming the currently observed planetary distribution
	as $0.33_{-0.25}^{+0.58}$ and $0.16_{-0.11}^{+0.29}$ for Hyades and Upper Scorpius,}
	respectively in Section \ref{sec:detectability_current}.
	{ Through the same calculations above, the normalized planet frequencies are derived as
	$0.079_{-0.079}^{+0.061}$ and $0.027_{-0.027}^{+0.019}$.}
	This result also suggests that the true planet frequency in Upper Scorpius
	is significantly lower than in Hyades.
	}

	In order to confirm these speculations,
	more samples of young planetary systems including other clusters are required.
	If future planet explorations at NIR wavelengths are performed,
	the detection efficiency will be enhanced to
	{ $0.19 \pm 0.04$, $0.15 \pm 0.04$, $0.11 \pm 0.03$, and $0.19 \pm 0.04$ }
	and the systems with the age of $\sim$ 100 Myr may be newly discovered
	based on the results in Section \ref{sec:nir}.
	\begin{figure}[t]
			\centering
			\includegraphics[width=9cm]{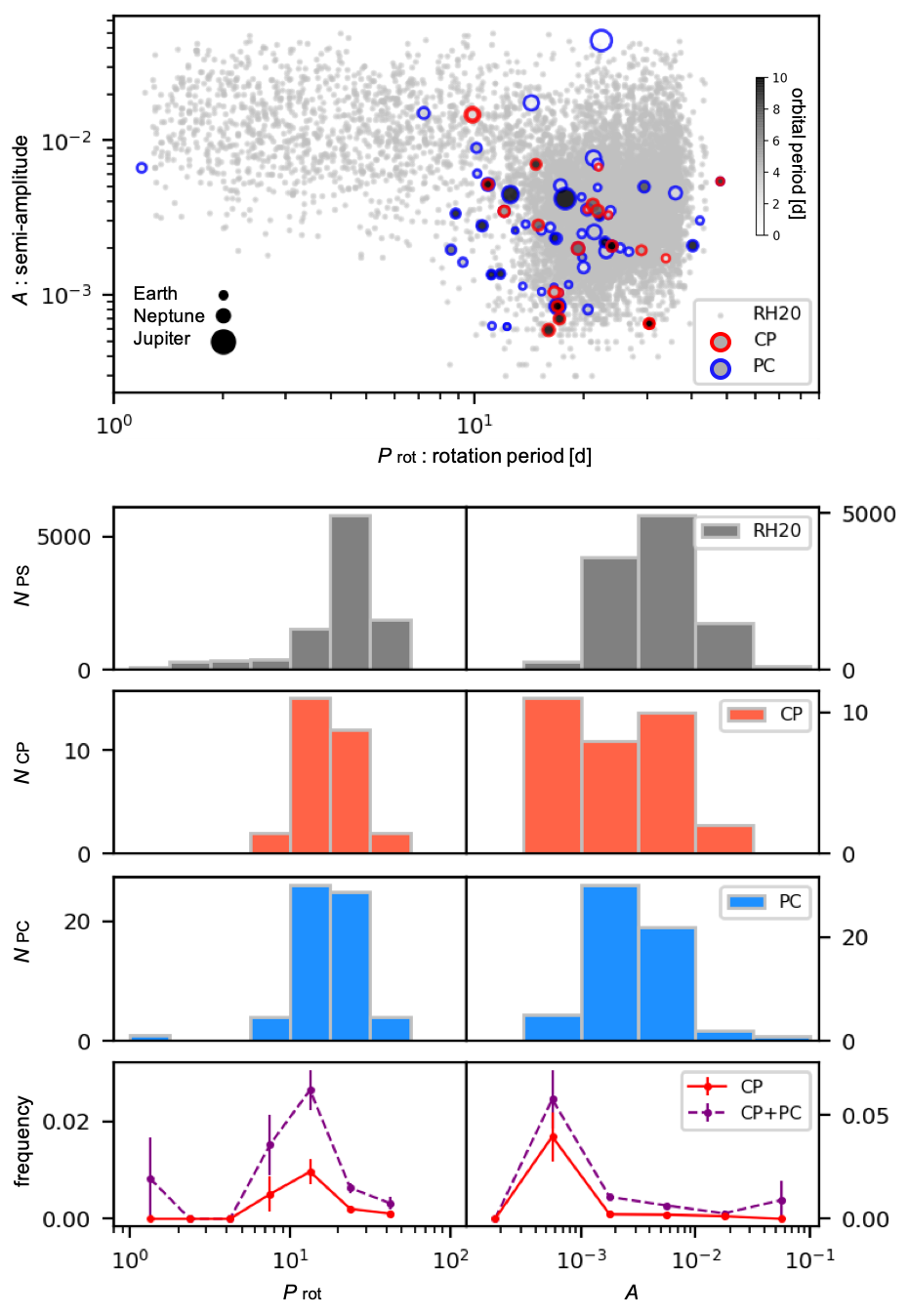}
			\caption{Distributions of periodic stars, planetary systems, and planetary candidates in $K2$
			with respect to the stellar rotation period (left column) and semi-amplitude of the light curve
			(right column).
			Top to bottom;
			scatter plot of $A$ versus $P_{\rm rot}$;
			histograms of all periodic stars, confirmed planets (CP), and
			planetary candidates (PC);
			line graphs showing the frequencies of CP and CP + PC among all periodic stars,
			respectively.
			In the top panel, the sizes and colors
			to the sizes and orbital periods, respectively, of the targets flagged as CPs (red) or PCs (blue).
			}
			\label{pic:current}
	\end{figure}

	\subsection{Implication to Current Planetary Systems}
	To understand the detection biases caused by stellar activity,
	we interpreted the current planetary distribution.
	We first constructed a scatter plot of $P_{\rm rot}$ vs. $A$ (top panel of Figure \ref{pic:current}).
	Each gray dot is a periodic star (PS; i.e., a star showing photometric periodicity induced by the rotation)
	with an effective temperature
 	between 3000 K and 4000 K in Campaigns 0 - 18 of the $K2$ mission
	measured in \citet{Reinhold2020}
	and each of these PSs may or may not host a planetary system.
	{ We also removed the stars belonging to typical clusters in {\it K2}, Hyades, Praesepe, Pleiades,
	Upper Scorpius, and $\gamma$ Ophiuchi,
	referring the EPIC list in the literatures \citep{Rebull2016, Douglas2017, Rebull2018, Douglas2019}
	to focus on the matured planetary systems.}
	The samples of \citet{Reinhold2020}
	require a match among the LS, the autocorrelation \citep{McQuillan2013},
	and the wavelet analyses for determining their rotation periods \citep{Torrence1998}.
	The red- and blue-edged circles in the top panel of Figure \ref{pic:current} are
	the targets flagged as confirmed planets (CPs)
	and planetary candidates (PCs) in the \citet{ExoFOPK2}\footnote{https://exofop.ipac.caltech.edu/k2/},
	{ where young CPs and PCs are removed as in the case of the PS.}
	{ To obtain the rotation periods of the CPs and PCs,
	we ran the LS periodogram after masking the transit and eclipse signals
	and detrending as described in Section \ref{sec:injection} with the K2SFF light curves.
	We determined the first independent peak as the rotation period for each target.}
	The vertical axis of the graph is the semi-amplitude of the 5th - 95th photometric variation,
	equivalent to $R_{\rm var}/2$ in \citet{Reinhold2020} for the gray points.
	The middle three rows of Figure \ref{pic:current} histograms of PSs, CPs, and PCs
	with respect to $P_{\rm rot}$ (left panels) and $A$ (right panels);
	the bottom panels, plot the frequencies of CP and CP+PC to the PS
	[$N_{\rm CP}/N_{\rm PS}$ and $(N_{\rm CP} + N_{\rm PC})/ N_{\rm PS}$, respectively]
	in each bin of the preceding histograms.
	The error values are calculated based on the binomial distribution,
	but are likely underestimated due to the small numbers of the CP and PC samples.

	In the scatter plot, the CPs and PCs are distributed
	around $P_{\rm rot}$ $\sim 10$ and $A$ $\sim 10^{-3}$.
	The CP targets with long orbital periods
	tend to distribute around the lower right region of the plot.
	No CPs are found at { rotation} periods below five days.
	The size is positively correlated with $A$ among the PCs (less so among the CPs).
	{ The orbital photometry of Jupiter-sized targets can contaminate the stellar activity
	and make $A$ larger because such targets are often detected as
	false positives (FP) of eclipsing binaries \citep[e.g., KOI-977; ][]{Hirano2015}.}
	One PC (EPIC 20192810.01) locates among the shortest $P_{\rm rot}$,
	but after visual inspection of the light curve,
	we concluded that EPIC 21092810.01 is an FP of stellar rotational activity.

	The distributions of $N_{\rm CP}$ and $N_{\rm PC}$ are slightly shifted
	towards shorter $P_{\rm rot}$ than $N_{\rm PS}$.
	Consequently, the CPs and PCs are most commonly found from approximately 7 to 20 days,
	while this systematic feature is small among the CPs alone.
	Meanwhile, the CPs show a wider distribution of semi-amplitudes of the light curves $A$ than the PSs.
	The limited number of PSs in the low $A$ region might be attributed to the high requirement
	of the period determination in \citet{Reinhold2020}.
	The occurrence frequencies of CP and PC appear to be high around low-activity stars
	with $A$ less than $10^{-3}$.

	Compared with the systematic features above,
	the recovery rate in Figure \ref{pic:number_of_samples}
	moderately vary as functions of $P_{\rm rot}$ and $A$.
	For the stellar rotation, while a peak locates at approximately 10 days
	in the frequencies of CPs and CPs + PCs (left bottom in Figure \ref{pic:current}),
	the decrease of recovery rate between about 3 and 10 days is almost $20\%$
	when the orbital period is 10 d (left bottom of $P_{\rm orb} = 10$ d in Figure \ref{pic:number_of_samples}).
	The frequency of semi-amplitudes of the light curves rises suddenly to its maximum at $A \approx 10^{-3}$
	(right bottom in Figure \ref{pic:current}),
	{ but the decrease in detectability is moderate }
	by around $60\%$ from $A = 10^{-4}$ to $10^{-1}$
	(right bottom for each $P_{\rm rot}$ in Figure \ref{pic:number_of_samples}).
	{ These frequency trends in Figure \ref{pic:current} indicate that
	a true planetary radius distribution rapidly changes
	while the detection efficiency gently changes with $P_{\rm rot}$ and $A$.}
	Especially for $A$,
	many CPs and PCs are missing in the high amplitude region regardless of the selection biases
	of \citet{Reinhold2020}.

	Our results on the detectability of transiting signals
	simply suggest a low planet frequency around active cool stars { intrinsically}
	in the current observational data.
	This idea was originally proposed by \citet{McQuillan2013},
	who analyzed the first {\it Kepler} observational results.
	Rapid rotators in low-mass stars in matured age can be explained by
	decrease of the magnetic braking effect in the PMS phase
	and retention of rapid rotators after the ZAMS of M-dwarfs \citep{Reiners2012},
	spin acceleration due to transportation of angular momentum
	by planetary migration or engulfment in host stars \citep{Bolmont2012, Gallet2019},
	or protoplanetary disk dissipation in the early stage due to the binarity of low-mass stars,
	which does not affect stellar rotation \citep{Stauffer2018}.
	To discuss relationships between the above scenarios and the lack of planets around rapidly rotators,
	further investigations of the planetary frequency in young stage are required.

\section{Summary}\label{sec:sum}

	We investigated the detectability of planetary candidates
	in the presence of photometric variations typically shown by young cool stars.
	The transit detectability was evaluated by
 	{ making the SDE maps} of the {\it K2} light curves
	collected in the four clusters --- Hyades, Praesepe, Pleiades, and Upper Scorpius ---
	using the planetary transit model over a wide range of orbital periods and planetary radii.
	The systematic trends of the transit detectability were similar in each cluster.
	We concluded that the lack of planets in Pleiades \citep{Gaidos2017}
	is likely due to rapid rotations of M-dwarfs around the ZAMS.

	In addition, we showed that the detectability of young planets is improved
	by future photometric observations in the NIR, such as by the JASMINE mission \citep{Gouda2018}.
	For the targets with rotational periods around one day or smaller,
	the detectability at a threshold of $1\%$ transit depth was
	(at most) approximately $20\%$  higher at NIR wavelengths than at visible wavelengths.
	This improvement suggests that
	future NIR photometric monitoring will find planetary systems that were overlooked
	in the {\it K2} or {\it TESS} missions
	and will constrain the scenarios of planetary and stellar activity evolution.

	Comparing the recovery rate to the current planet frequency as functions of stellar activity,
	confirmed planets and candidates around M-dwarfs
	in the {\it K2} mission are likely missing around active stars.
	Whereas more observational researches are required,
	this biased distribution may be an evidence that planet formation and/or evolution are prevented around
	rapid rotators in M-dwarfs.
	Follow-up observations at NIR wavelengths are expected to impose further constraints on
	roles of stellar activity in young planetary systems.
	\begin{figure}[t]
			\centering
			\includegraphics[width=8cm]{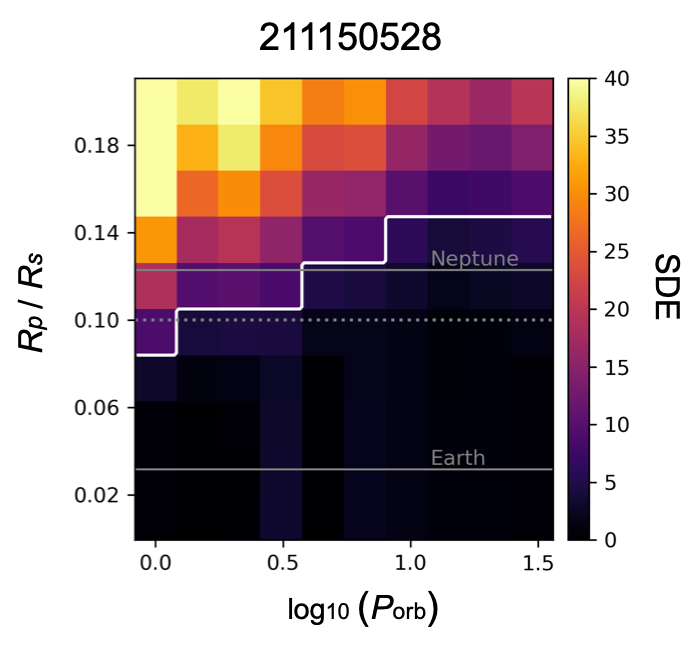}
			\caption{The SDE map for EPIC 211150528 which is a rapidly rotating target
			for comparison with lower panels of Figure 7 in \citet{Rizzuto2017}.}
			\label{pic:211150528}
	\end{figure}
\acknowledgments
	We would like to thank Dr. Hajime Kawahara for fruitful discussions.
	This work was supported by Japan Society for Promotion of Science (JSPS) KAKENHI
	Grant Numbers JP19J21733,  JP19K14783, and JP21H00035. 
	{ EG is supported by NASA Award 80NSSC20K0957
	(Exoplanets Research Program) and NSF Award 1817215 (Astronomy \& Astrophysics Research Grants).}
 	This work has made use of data from the European Space Agency (ESA) mission
	Gaia (\url{https://www.cosmos.esa.int/gaia}), processed by the Gaia
	Data Processing and Analysis Consortium (DPAC,
	\url{https://www.cosmos.esa.int/web/gaia/dpac/consortium}). Funding for the DPAC
	has been provided by national institutions, in particular the institutions
	participating in the Gaia Multilateral Agreement.

\appendix
	\section{Uncertainty in the Estimation of Stellar Properties}\label{ap:iso}
	\begin{figure}[t]
		\centering
		\includegraphics[width=8cm]{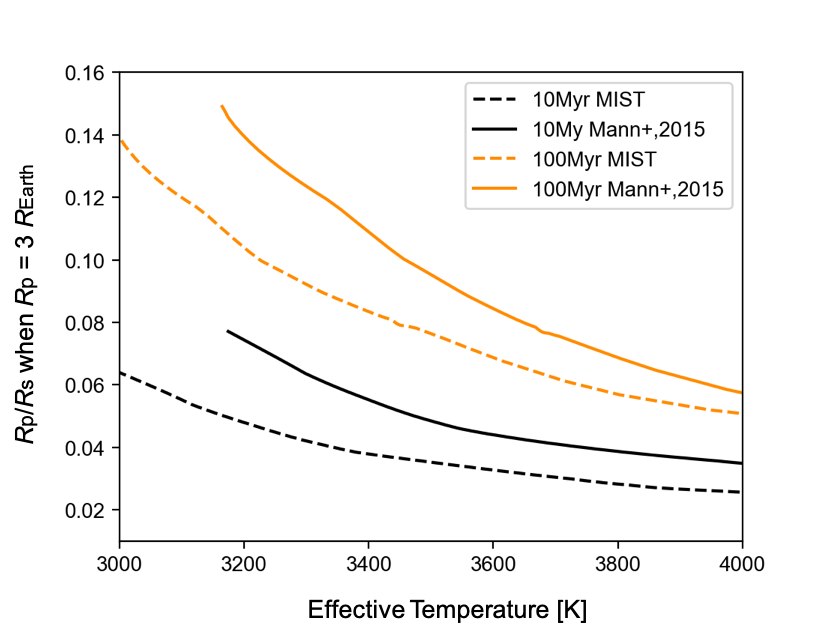}
		\caption{ $T_{\rm eff}$ vs. $R_{\rm p}/R_{\rm s}$ plot when $R_{\rm p} = 3 R_{\oplus}$
		with \citet{Mann2015} (solid)
		and the MIST \citep{Dotter2016, Choi2016} (dashed) ; the black
		and orange lines are 10 Myr and 100 Myr, respectively.
		}
		\label{pic:compare_iso}
	\end{figure}

	{
	We discuss the uncertainty in the stellar properties estimated
	with the empirical formula in \citet{Mann2015} (see Section \ref{sec:target} and \ref{sec:injection}),
	which is required since stellar radius in the PMS phase (10 - 100 Myr) is likely inflated.
	We calculate the $T_{\rm eff}$ vs. $R_{\rm s}$ plot
	using the MIST stellar isochrones \citep{Dotter2016, Choi2016}
	with ages of 10 Myr and 100 Myr which correspond to the Upper Scorpius and Pleiades ages, respectively.
	The stellar radius is derived via the Stefan-Boltzmann law with the effective temperature and luminosity.
	To clarify how much effect the uncertainty has on our SDE estimations,
	we converted $R_{\rm s}$ into $R_{\rm p}/R_{\rm s}$ when $R_{\rm p} = 3 R_{\oplus}$
	which corresponds to typical value of the vertical axis in Figure \ref{pic:cmap},
	and show the result in Figure \ref{pic:compare_iso}.
	Our estimations with \citet{Mann2015} seem to be larger by about 0.02 in $R_{\rm p}/R_{\rm s}$
	than with the MIST for both 10 Myr and 100 Myr,
	and this difference corresponds to about one pixel grid in the SDE maps.
	Therefore, we presumably overestimate the planet detectability in the Upper Scorpius and Pleiades
	by approximately 10 \%,
	although this level of systematic error does not affect the overall results
	and conclusion discussed in the text.
	}

	\section{Comparison to Other Detrending Method}\label{ap:detrend}
\begin{figure}[t]
			\centering
			\includegraphics[width=16cm]{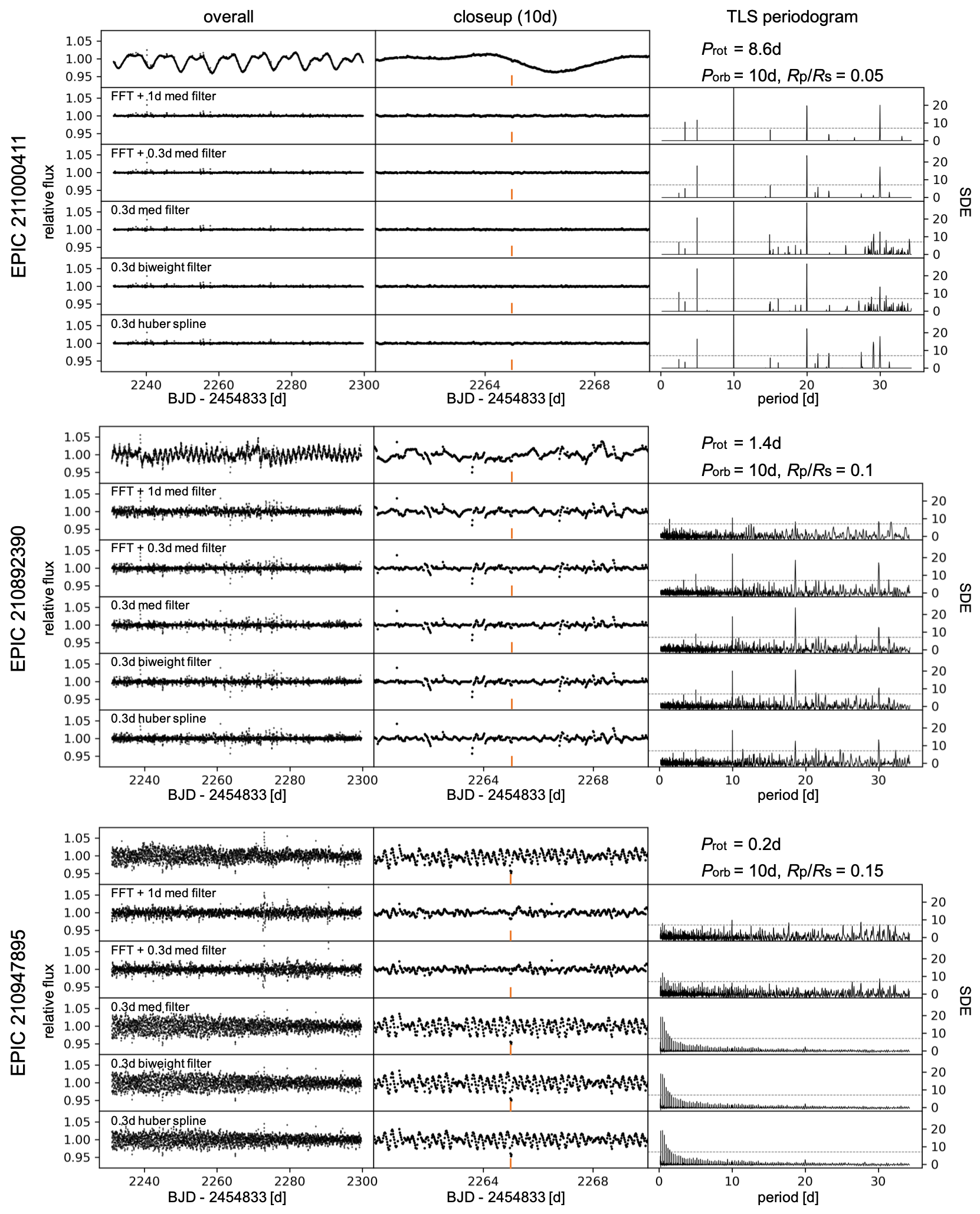}
			\caption{ Light curves and TLS results for EPIC 211000411, EPIC 210892390,
			and EPIC 210947895.
			For each target, from the top to the bottom,
			light curve after subtraction of long-term modulation and
			light curve detrended with the FFT + 1 d median filtering,
			the FFT + 0.3 d median filtering, only the 0.3 d median filtering,
			the 0.3 d biweight filtering and the 0.3 Huber spline detrending;
			from the left to the right, overall and colseup view of light curves and the TLS periodogram.
			In the middle column, we indicate transit signals with orange line;
			in the right column, dashed horizontal line represents the SDE of 7.
			}
			\label{pic:detrend}
	\end{figure}

	{
	We discuss the efficiency of detrending method introduced in Section \ref{sec:recovery}
	comparing to other approaches.
	Firstly, \citet{Rizzuto2017} have developed two transit recovery algorithms
	called notch filter pipeline and Locally Optimized Combination of Rotation (LOCoR).
	The first one performs fitting a transit-shaped notch and quadratic baseline in a 0.5 or 1 d window
	to the stellar activity and transit dips simultaneously,
	and has sensitivity to targets with $P_{\rm rot} \geq 1 -1.5$ d.
	The latter one models stellar variability by a linear combination of observed rotation periods for each target
	and is effective for the most rapid rotators with $P_{\rm rot} \leq 2$ d.
	As reference, \citet{Rizzuto2017} show results of their injection recovery test
	using both of these methods in their Figure 7 for EPIC 210892390 and EPIC 211150528.
	When focusing on better ones,
	detection borders lie in about 3 - 4 $R_{\oplus}$ for the notch filter to EPIC 210892390
	and about 2.5 - 3.5 $R_{\oplus}$ for the LOCoR to EPIC 211150528.
	For comparison, we also show the SDE maps for EPIC 210892390 and EPIC 211150528
	in Figure \ref{pic:cmap} and Figure \ref{pic:211150528}, respectively.
	The boundaries exist in approximately 3.5 - 5 $R_{\oplus}$ and 2.5 - 4 $R_{\oplus}$
	for EPIC 210892390 and EPIC 211150528, respectively,
	while we show them with planetary radii in $R_{\rm p}/R_{\rm s}$ and $P_{\rm rot}$ in log scale.
	These differences seem to be corresponding to a few pixels and negligible for the discussions.

	Next, we discuss the efficiency of some detrending methods before the TLS \citep{Hippke2019}.
	We tested the following five ways to detrend light curves for the three target with different rotation periods,
	EPIC 211000411, EPIC 210892390, and EPIC 210947895.
	\begin{enumerate}
	\item FFT filter + 1 d - median filter (performed in Section \ref{sec:recovery}.)
	\item FFT filter + 0.3 d - median filter
	\item 0.3 d - median filter
	\item 0.3 d - Tukey's biweight filter \citep{Tukey1977}
	\item 0.3 d - Huber spline detrend \citep{Huber1964}
	\end{enumerate}
	The latter two methods which are based on the robust statistics
	are evaluated as ideal methods in \citet{Hippke2019c}.
	We employed \texttt{W$\bar{\texttt o}$tan} \citep{Hippke2019c},
	a comprehensive time-series detrending package implemented in \texttt{Python}
	for the biweight and Huber spline analyses.
	We injected a planet with the orbital $P_{\rm orb}$ of 10 d
	and $R_{\rm p}/R_{\rm s}$ of 0.05, 0.1, 0.15
	for EPIC 211000411, EPIC 210892390, and EPIC 21097895, respectively.
	These planetary radii are selected around the detection boundary.

	The results are shown in Figure \ref{pic:detrend}.
	For each target, the top panel shows a light curve after subtraction of long-term modulation.
	From the second panel to the bottom, we display detrended light curves with the five methods above.
	We show total light curves, closeup view with 10 d - window, and the TLS periodogram
	from left to right.
	The horizontal dashed line in the TLS periodogram indicates detection threshold with SDE of 7.
	For EPIC 211000411 whose rotation period is 8.6 d,
	there is no significant differences around the first peak among the five methods.
	This is due to that photometric variations longer than a few days
	can be easily removed by 1 d - window scaled detrending.
	For EPIC 210892390 with the rotation period of 1.4 d,
	our FFT + 1d median shows lower SDE than the other methods,
	while the detection is significant.
	Because the periodogram of the FFT + 0.3 d median shows comparable SDE
	to that of the 0.3 d biweight and huber spline,
	this decline likely depends on the window size.
	Also, the red noise is predominant in the original light curve,
	it is not possibly caused by the rotational moduration.
	For EPIC 210947895, the most rapidly rotator with 0.2 d,
	only our approach successfully detected the transit signal.
	This indicates that the rotational noise could be removed in the FFT filtering
	and short window sized filtering may degenerate the red noise and transit signal.
	Thus, we conclude that our approach, the FFT + 1 d median,
	is likely more sensitive to extremely rapid rotators than other detrending methods
	and is appropriate when we discuss the lack of the planets around active stars.
	Note that discussion here is specific to the {\it K2} light curves
	which contain about 50 data points in a 1 d sized bin.
	Because the {\it TESS} has a few hundreds of data points in similar time scale,
	the results will be different.
	}

	\section{Validity of the 10 $\times$ 10 SDE map}\label{ap:3030}
	\begin{figure}[t]
			\centering
			\includegraphics[width=18cm]{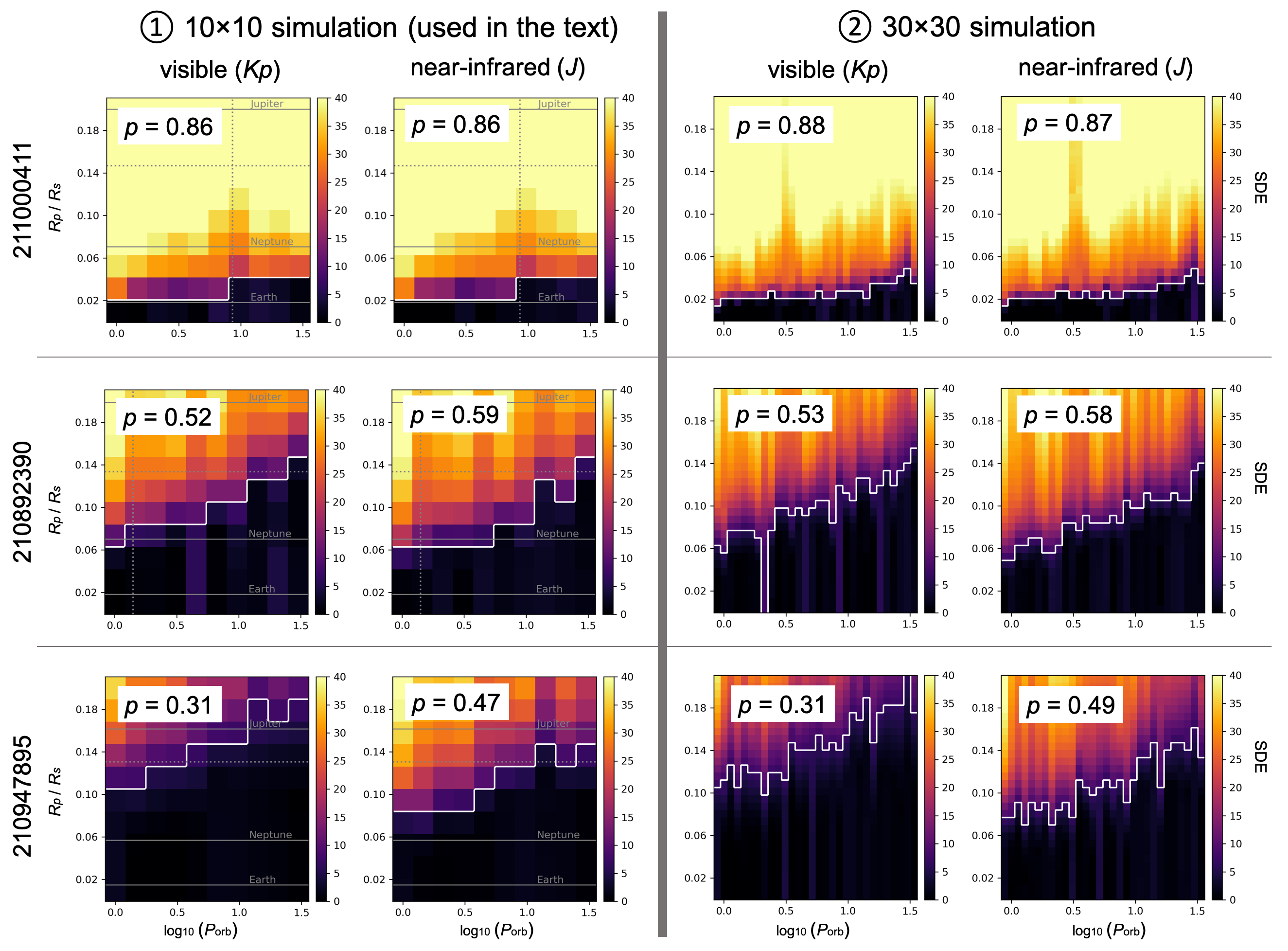}
			\caption{Comparisons of results between the 10 $\times$ 10 (Figure \ref{pic:cmap})
			and 30 $\times$ 30 simulations for EPIC 211000411, EPIC 210892390, and EPIC 210947895.
			}
			\label{pic:3030}
	\end{figure}
	{To ensure the validity of the 10 $\times$ 10 SDE maps in Section \ref{sec:injection},
	we show a comparison for the typical three targets
	with simulation results with 30 $\times$ 30 grids in Figure \ref{pic:3030}.
	The left and right panels are the results of the 10 $\times$ 10
	and 30 $\times$ 30 simulations, respectively.
	We show the recovery rate $p$ (see Section \ref{sec:totaltrend}) in the left top of each panel.
	The $p$ almost agree within about 1\% error, and features of the colormaps
	seem to be consistent between the two simulations.
	Thus, we conclude that the 10 \% 10 simulations are sufficient for this study.
	{ There are some spikes on the SDE distribution.
	For example, $\log{P_{\rm orb}} = 0.5$ in EPIC 211000411
	and $\log{P_{\rm orb}} = 0.3$ in EPIC 210892390 show systematically high values along the vertical axis.
	This might be caused by mis-detection of the transit signal in the TLS algorithm,
	because periodic residuals remain after the FFT filtering.}
	}

 \bibliographystyle{aasjournal}
 \bibliography{miyakawa21.bib}
\end{document}